\documentclass[11pt]{amsart}

\usepackage{amsmath}
\usepackage{blkarray}
\usepackage{subfig}
\usepackage{graphicx}
\usepackage{multirow}
\usepackage{amssymb}
\usepackage{bbm}
\usepackage{amsthm}
\usepackage{color}
\usepackage{tikz}

\usetikzlibrary{automata,positioning}

\newtheorem{theo}{Theorem}
\newtheorem{prop}{Proposition}
\newtheorem{lem}{Lemma}
\newtheorem{cor}{Corollary}
\newtheorem{Rem}{Remark}

\allowdisplaybreaks

\begin{document}

\title[A new class of conditional Markov jump processes]{A new class of conditional Markov jump processes with regime switching and path dependence: properties and maximum likelihood estimation }
\author{ Budhi Surya}

\address{Budhi Surya:
School of Mathematics and Statistics, Victoria University of Wellington, Kelburn PDE Gate 7, Wellington 6140, New Zealand; Email address: budhi.surya@msor.vuw.ac.nz.}



\begin{abstract}
This paper develops a new class of conditional Markov jump processes with regime switching and paths dependence. The key novel feature of the developed process lies on its ability to switch the transition rate as it moves from one state to another with switching probability depending on the current state and time of the process as well as its past trajectories. As such, the transition from current state to another depends on the holding time of the process in the state. Distributional properties of the process are given explicitly in terms of the speed regimes represented by a finite number of different transition matrices, the probabilities of selecting regime membership within each state, and past realization of the process. In particular, it has distributional equivalent stochastic representation with a general mixture of Markov jump processes introduced in Frydman and Surya \cite{Frydman2020}. Maximum likelihood estimates (MLE) of the distribution parameters of the process are derived in closed form. The estimation is done iteratively using the EM algorithm. Akaike information criterion is used to assess the goodness-of-fit of the selected model. An explicit observed Fisher information matrix of the MLE is derived for the calculation of standard errors of the MLE. The information matrix takes on a simplified form of the general matrix formula of Louis \cite{Louis}. Large sample properties of the MLE are presented. In particular, the covariance matrix for the MLE of transition rates is equal to the Cram\'er-Rao lower bound, and is less for the MLE of regime membership. The simulation study confirms these findings and shows that the parameter estimates are accurate, consistent, and have asymptotic normality as the sample size increases.
\end{abstract}

\maketitle

\section{Introduction}
Markov jump process has been one of the most important probabilistic model since its introduction by Markov \cite{Markov06} in 1906. See Seneta \cite{Seneta} for history of the creation of the Markov chains. It is a simple parametric model, but yet a rather complex nonparametric model for describing sequence of events where development of a future event depends only on the current state and the duration it takes. The distribution parameter of the process is given in terms of a transition matrix $Q$ whose $(x,y)-$component $q_{xy}$ defines the rates of making a jump from a state $x$ to another state $y$. 

Statistical inference of the transition matrix was discussed in Albert \cite{Albert}. Distributional properties of the Markov process have been well studied in literature. We refer among others to Norris \cite{Norris}, Pardoux \cite{Pardoux}, Resnick \cite{Resnick}, and Stroock \cite{Stroock}. The Markov process has been widely used in variety of applications across various fields such as, among others, in modeling vegetation dynamics (Balzter \cite{Balzter}); demography (Nowak \cite{Nowak}); in marketing to model consumer relationship (Berger and Nasr \cite{Berger} and Pfeifer and Carraway \cite{Pfeifer}), and to identify substitutions behavior of customers in assortment problem (Blanchet et al. \cite{Blanchet}); in describing credit rating transitions used in many credit risk and pricing applications (Jarrow and Turnbull \cite{Jarrow1995}, Jarrow et al. \cite{Jarrow1997}, Bielecki et al. \cite{Bielecki}, and Bielecki and Rutkowski \cite{Bielecki2002}); in queueing networks and performance engineering (Bolch et al. \cite{Bolch}, Pardoux \cite{Pardoux}). It has also been widely used for multi-state analysis of life history and panel data in which a Markov jump process is used to describe the progression of health status of a cohort. See for e.g., Aalen and Gjessing \cite{Aalen2001}, and Cook and Lawless \cite{Cook2018}. Statistical softwares for the multi-state analysis by de Wreede et al. \cite{Wreede} and Jackson \cite{Jackson} are freely available in an R package which make implementation of the Markov model against life history and panel data possible and convenient. 

Due to its lack of memory property, future evolution of the Markov jump process depends only on its current state, but not on its past history. Furthermore, transition from one state to any phase of the state space is only determined by duration of the transition, but not on the current time (''age'' of the process). These might suggest a needed extension of the Markov process. 

In recent years there have been attempts to extend the Markov model into one which relaxes the memoryless and stationary properties. Bielecki et. al. \cite{Bielecki2017}, and Jakubowski and Nieweglowski \cite{Jakubowski} proposed an $\mathbb{F}-$conditional doubly stochastic Markov chains given natural filtration (past realization of the sample paths) of the Markov jump process and the filtration generated by exogenous covariate defined by a stochastic intensity as such that future development of a state of the process depends on external factor in non-stationary way. However, the constructed process retains the memoryless property and did not address a possible heterogeneity of cohort in its application using life history or panel data. No details on a maximum likelihood estimation of the model was provided. In 1955, Blumen et al. \cite{Blumen} introduced the seminal \textit{mover-stayer} (MS) model, which is the mixture of two discrete-time Markov chains used to account for heterogeneity in a cohort of workers in labor market. It was the first mixture of Markov chains considered in the literature. For its continuous-time counterpart, the MS model describes a cluster of workers consisting of stayers (workers who always stay in the same job category) and movers (workers who move to other job category according to a Markov process). Frydman \cite{Frydman} discussed maximum likelihood estimation (MLE) of the MS model by direct maximization of the observed likelihood function. The EM algorithm for the estimation of the MS model was proposed by Fuchs and Greenhouse \cite{Fuchs}. 

Frydman \cite{Frydman2005} extended the MS model \cite{Blumen} into a particular mixture of Markov jump processes. In its implementation using life history data, the model allows each cohort to move with respect to its own Markov process. The Markov mixture model \cite{Frydman2005} was successfully applied to modeling credit ratings migration by Frydman and Schuermann \cite{Frydman2008} in which they found that bonds of the same credit rating may move at different speeds to other credit ratings. In addition to that, they observed that the inclusion of past credit ratings improves out-of-sample prediction of Nelson-Aalen estimate of credit default intensity. The empirical findings in \cite{Frydman2008} suggest that the credit rating migration could be represented by a mixture of Markov jump processes moving at different speeds. The mixture model \cite{Frydman} was also successfully applied to clustering of categorical time series by Pamminger and Fruhwirth-Schnatter \cite{Pamminger}. Surya \cite{Surya2018} revisited \cite{Frydman2008} and gave further explicit distributional identities of the mixture process, in particular in the presence of an absorbing state. Despite each underlying Markov process of the mixture \cite{Frydman2005} has different exit rates, the respective embedded Markov chains of the underlying Markov processes have the same transition matrix. This constraint on the transition matrix of the embedded Markov chains may obscure clustering of an heterogeneous cohort, which might result in being rejected using the likelihood ratio (LR) test when compared to a general mixture of Markov jump processes. See Frydman and Surya \cite{Frydman2020} for details of the LR test, MLE of the general mixture, and asymptotic properties of the MLE. And to Surya \cite{Surya2020} for the joint probability distributions of absorption times of the mixture process.

By formulating the general mixture process \cite{Frydman2020} and \cite{Surya2020} in terms of its underlying Markov jump processes, a distributional equivalent stochastic representation of the mixture process is derived in terms of a \textit{regime switching conditional Markov jump process with path dependence}. The governing equations for the dynamics of the conditional Markov jump process (CMJP) are presented in terms of an increasing sequence of epoch times of the process and the states reached at the epoch times. 

This paper is organized as follows. Section \ref{sec:sec2} outlines the research contribution of the paper. Representation of the general mixture process \cite{Frydman2020}, \cite{Surya2020} and the proposed conditional Markov jump process are discussed in details in this section. Section \ref{sec:sec3} presents distributional properties of CMJP. Maximum likelihood estimates of the distribution parameters and EM algorithm for the estimation are discussed in Section \ref{sec:sec4}. Also, the observed Fisher information matrix of the estimates is derived in this section. Large sample properties of the MLE are presented in Section \ref{sec:sec5}. To verify the main results presented in Sections \ref{sec:sec2}-\ref{sec:sec5}, a series of simulation studies are performed in Section \ref{sec:sec6}. The simulation study shows that the parameter estimates are accurate with their biases and root mean square errors decreasing by the size of generated sample paths of CMJP. Akaike information criterion is employed to test goodness-of-fit of the selected model against the data it was generated from. Kolmogorov-Smirnov test is conducted for each parameter bias. The test is found to be statistically significant at the acceptance level $\alpha=5\%$ for each parameter bias, confirming the consistency and asymptotic normality of the estimates. Section \ref{sec:sec7} concludes this paper.

\section{Research contribution}\label{sec:sec2}
Consider $M$ independent finite-state right-continuous Markov jump processes $X^{(1)},X^{(2)},\cdots,X^{(M)}$ defined on the same finite-state space $\mathbb{S}=\{1,\cdots,p\}$ with the same state transition diagram. Each Markov process $X^{(m)}=\{X^{(m)}(t): t\geq 0\}$, for $m=1,\cdots,M$, makes $x\rightarrow y$ transitions w.r.t. its own intensity matrix $Q_m=[q_{xy,m}]_{xy}$, with $(x,y)\in \mathbb{S}$, satisfying the following condition:
\begin{eqnarray*}
q_{xx,m}<0, \quad q_{xy,m}\geq 0, \;\; y\neq x, \;\; \textrm{s.t.} \;\; q_{xx,m}=-\sum\limits_{y\neq x,y\in\mathbb{S}} q_{xy,m}.
\end{eqnarray*} 
The intensity $q_{xy,m}$ defines the transition rate of making an instantaneous jump from states $x$ to $y$, with $(x,y)\in \mathbb{S}$, under the Markov process $X^{(m)}$, defined for each $1\leq m\leq M$ by
\begin{align}\label{eq:rate}
q_{xy,m}=\lim_{h\downarrow 0} \frac{1}{h}\Big(\mathbb{P}\big\{X(h)=y \big\vert \Phi=m, X(0)=x\big\} - \delta_x(y)\Big),
\end{align}
where $X$ represents the observed process, whilst $\Phi$ specifies the underlying dynamics $X^{(m)}$ of $X$, see (\ref{eq:mixture}) for construction. The function $\delta_x(y)$ has value one if $y=x$ or zero otherwise defined by 
\begin{align*}
\delta_x(y)=
\begin{cases}
1, & \textrm{if $y=x$},\\
0, &\textrm{otherwise}.
\end{cases}
\end{align*}
Note that each intensity matrix $Q_m$, with $1\leq m\leq M$, can be rewritten in terms of a transition matrix $\Pi_m=[\pi_{xy,m}]_{xy}$ of the embedded Markov chain of $X^{(m)}$ and an identity matrix $\mathbf{I}$ as
\begin{align}\label{eq:intQ}
Q_m=-\textrm{diag}\big(q_{11,m},\cdots,q_{pp,m}\big)\big(\Pi_m-\mathbf{I}\big),
\end{align}
where the $(x,y)-$element $\pi_{xy,m}$ of the transition matrix $\Pi_m$ is the defined by
\begin{align*}
\pi_{xy,m}=
\begin{cases}
\frac{q_{xy,m}}{q_{x,m}}, & y\neq x\\[5pt]
0, & y=x.
\end{cases}
\end{align*}

\begin{figure}[t!]
\begin{center}
  \begin{tikzpicture}[font=\sffamily]

        \tikzset{node style/.style={state,
                                    minimum width=1cm,
                                    line width=0.75mm,
                                    fill=white!20!white}}

           \tikzset{My Rectangle2/.style={rectangle,draw=brown,  fill=yellow, thick,
    prefix after command= {\pgfextra{\tikzset{every label/.style={blue}}, label=below}}
    }
}

          \tikzset{My Rectangle3/.style={rectangle, draw=brown, fill=white, thick,
    prefix after command= {\pgfextra{\tikzset{every label/.style={blue}}, label=below}}
    }
}

        \node[node style] at (3, 3.75)     (s11)     {$1$};
         \node[draw=none, label={}] at (3, 1.875)     ({})     {$\vdots$};
          \node[node style] at (7, 3.75)     (s12)     {$x$};
         \node[node style] at (11, 3.75)    (s13)     {${p}$};
          \node[draw=none, label={}] at (11, 1.875)     ({})     {$\vdots$};

          \node [My Rectangle3, label={} ]  at (-1, -1)     (s0)     {$\mathbf{p}$};

          \node[node style] at (3, 0)     (s21)     {$1$};
        \node[node style] at (7, 0)     (s22)     {$x$};
         \node[node style] at (11, 0)    (s23)     {${p}$};
 
        \node[node style] at (3, -3.75)     (s31)     {$1$};
         \node[draw=none, label={}] at (3, -1.875)     ({})     {$\vdots$};
        \node[node style] at (7, -3.75)     (s32)     {$x$};
         \node[node style] at (11, -3.75)    (s33)     {${p}$};
          \node[draw=none, label={}] at (11, -1.875)     ({})     {$\vdots$};

      \node [My Rectangle3, label={}] at  ([shift={(-5em,0em)}]s11.west) (p10) {$(X^{(1)},\boldsymbol{\alpha}^{(1)})$};

      \node [My Rectangle3, label={}] at  ([shift={(-5em,0em)}]s21.west) (p20) {$(X^{(m)},\boldsymbol{\alpha}^{(m)})$};

     \node [My Rectangle3, label={}] at  ([shift={(-5em,0em)}]s31.west) (p30) {$(X^{(M)},\boldsymbol{\alpha}^{(M)})$};

        \draw[every loop,
              auto=right,
              line width=0.75mm,
              >=latex,
              draw=orange,
              fill=orange]

            (s11)  edge[bend right=20, auto=left]        node {$q_{1x,1}$} (s12)
            (s12)  edge[bend right=20, auto=left]                       node {$q_{x1,1}$}  (s11)

             (s13)  edge[bend right=20, auto=left]                      node {$q_{px,1}$}  (s12)
             (s12)  edge[bend right=20, auto=left]       node {$q_{xp,1}$}  (s13)

             (s13)  edge[bend right=25]                      node {$q_{p1,1}$}  (s11)
             (s11)  edge[bend right=25, auto=right]     node {$q_{1p,1}$}  (s13)


              (s21)  edge[bend right=20, auto=left]      node {$q_{1x,m}$} (s22)
              (s22)  edge[bend right=20, auto=left]                     node {$q_{x1,m}$}  (s21)

              (s23)  edge[bend right=20, auto=left]                     node {$q_{px,m}$}  (s22)
             (s22)  edge[bend right=20, auto=left]       node {$q_{xp,m}$}  (s23)

             (s23)  edge[bend right=25]                      node {$q_{p1,m}$}  (s21)
             (s21)  edge[bend right=25, auto=right]     node {$q_{1p,m}$}  (s23)


              (s31)  edge[bend right=20, auto=left]      node {$q_{1x,M}$} (s32)
              (s32)  edge[bend right=20, auto=left]                     node {$q_{x1,M}$}  (s31)

             (s33)  edge[bend right=20, auto=left]                      node {$q_{px,M}$}  (s32)
             (s32)  edge[bend right=20, auto=left]       node {$q_{xp,M}$}  (s33)

             (s33)  edge[bend right=25]                      node {$q_{p1,M}$}  (s31)
             (s31)  edge[bend right=25, auto=right]     node {$q_{1p,M}$}  (s33)
 

          (s0) edge [-,auto=left]  node {$p_1$} (s11)
           (s0) edge [-,auto=right]  node {$p_m$} (s21)
            (s0) edge [-,auto=right]  node {$p_M$} (s31);
 \end{tikzpicture}
 \caption{State diagram of the mixture process (\ref{eq:mixture}) with $\alpha_x^{(\bullet)}=\mathbb{P}\{X_0=x\vert \Phi=\bullet\}$.}\label{fig:mixture2}
\end{center}
\end{figure}
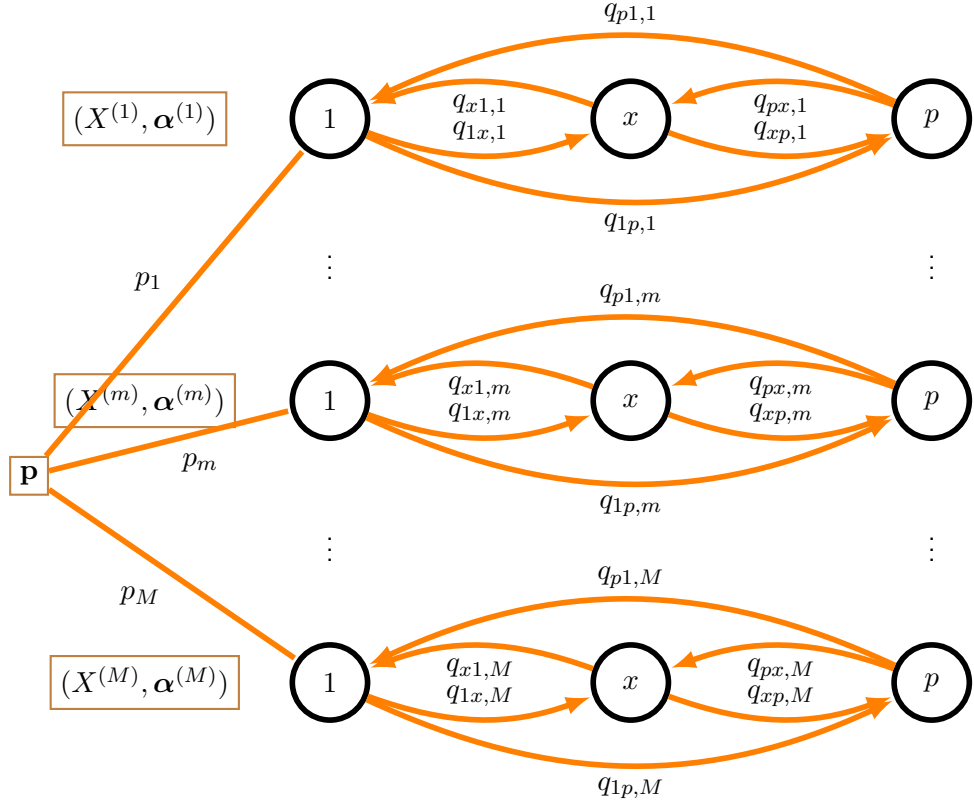

Let $\Phi$ be a categorical random variable, independent of $\{X^{(m)}\}$, having $M$ number of categories with distribution $p_m=\mathbb{P}\{\Phi=m\}$ s.t. $\sum_{m=1}^M p_m=1$. Denote by $\Phi^{(m)}=\mathbbm{1}_{\{\Phi=m\}}$ a Bernoulli indicator random variable with the probability of success $p_m$. Consider the mixture process
\begin{align}\label{eq:mixture}
X(t)=\sum_{m=1}^M \Phi^{(m)} X^{(m)}(t), \quad t\geq 0.
\end{align}
The state diagram of the mixture process $X$ (\ref{eq:mixture}) is illustrated in Figure \ref{fig:mixture2}. The diagram shows that conditional on the event $\{\Phi=m\}$, $X$ moves w.r.t the Markov process $X^{(m)}$, i.e., once the speed regime $\Phi$ is selected, $X$ evolves according to the selected Markov process. As $\sum_{m=1}^M \Phi^{(m)}=1$, $X$ defines a finite mixture of Markov processes $\{X^{(m)}:1\leq m\leq M\}$. Since the Markov processes $\{X^{(m)}\}$ are defined on the same state space $\mathbb{S}$ with the same transition diagram, there is uncertainty concerning which underlying process that drives the dynamics of $X$. That is, when $X$ is observed in a state $x\in\mathbb{S}$, there is one out of $M$ possible underlying processes $\{X^{(m)}\}$ by which $X$ arrived in $x$. Unlike its underlying processes, $X$ does not have the Markov property. This can be seen from the fact that $\Phi^{(m)}=1$ if and only if $X=X^{(m)}$, i.e., the observed process $X$ is driven by $X^{(m)}$. 

Since the regime membership $\Phi_m$ of the observed process is unobservable, it can be inferred from continuous observation $\mathcal{H}_t=\{X(u):0\leq u\leq t\}$ of the sample paths. Thus, the Bayesian update $\mathbb{P}\{\Phi=m\vert \mathcal{H}_t\}$ may depend on the sample paths of the observed process $X$. To be more precise,
\begin{eqnarray}\label{eq:phixt}
\phi_{x,m}(t):=\mathbb{P}\{\Phi=m \big\vert X_t=x, \mathcal{H}_{t-}\}=\frac{ \phi_{x_0,m} f_{\theta_m}\big(\mathcal{H}_{t,x}\big\vert \Phi=m\big)}{\sum_{\xi=1}^M \phi_{x_0,\xi} f_{\theta_{\xi}}\big(\mathcal{H}_{t,x}\big\vert \Phi=\xi\big)},
\end{eqnarray}
where $\phi_{x_0,m}=\mathbb{P}\{\Phi=m\vert X_0=x_0\}$ and $f_{\theta_m}\big(\mathcal{H}_{t,x}\big\vert \Phi=m)$ denotes the likelihood of observation $\mathcal{H}_{t,x}=\{X_t=x\}\cup \mathcal{H}_{t-}$ when $X$ moving w.r.t $X^{(m)}$ with transition intensity $\theta_m=\big(q_{xy,m}:(x,y)\in\mathbb{S}\big)$. Note that $\sum_{m=1}^M \phi_{x,m}(t)=1$ for every $x\in\mathbb{S}$, $t\geq 0$. From Theorem 3.1, p. 731, in Albert \cite{Albert},
\begin{align}\label{eq:likelihood}
f_{\theta_m}\big(\mathcal{H}_{t,x}\big\vert \Phi=m\big)= \alpha_{x_0}\prod_{k=1}^{n_t-1} \Big(q_{x_k,m} e^{-q_{x_k,m} (T_{k+1}-T_k)}\frac{q_{x_k x_{k+1},m}}{q_{x_k,m}}\Big) e^{-q_{x,m}(t-T_{n_t})},
\end{align}
with $q_{x,m}=-q_{xx,m}$, $\alpha_{x_0}=\mathbb{P}\{X_0=x_0\}$, whereas $n_t=\max\{k:T_k\leq t\}$ denotes the number of jumps before $t>0$, whilst $\{T_k\}_{k\geq 0}$ represents an increasing sequence of epoch times of $X$, i.e., $$T_{k+1}=\inf\{t>T_k: X(t)\neq x_k\}, \;\; \textrm{with} \;\; x_k=X(T_k),$$ with a specified initial time $T_0\geq 0$. Notice that the likelihood contribution of the sample paths on the last interval $[T_{n_t},t)$ is due to the right censoring of occupation time in the last state $x$. Some distributional properties of the mixture process $X$ (\ref{eq:mixture}) such as the transition matrix and, in particular, the joint probability distributions of absorption times were discussed in Surya \cite{Surya2018}. 

Based on the stochastic representation (\ref{eq:mixture}) of the mixture process $X$, the result below shows that $X$ is equivalent in distribution with a regime switching conditional Markov jump process (\ref{eq:main}).

\begin{theo}[\textbf{Regime switching conditional Markov jump process}]\label{theo:main}
The mixture of Markov jump processes (\ref{eq:mixture}) has equivalent distributional representation with $X=\{X(t):t\geq 0\}$ defined by
\begin{align}\label{eq:main}
X(t)=\sum_{n=0}^{\infty} X_n \mathbbm{1}_{[T_n,T_{n+1})}(t),
\end{align}
where $T_n$ and $X_n$, respectively the epoch time of $X$ and the observed state at $T_n$, satisfy for a given sequence of uniform random variables $(U_n,V_n,W_n)_{n\geq 0}$, independent of $X$, the recursive equations
\begin{align}
X_{n+1}=&F(X_n,\Phi_n,U_{n+1}),\label{eq:Xn}\\
T_{n+1}=&T_n-\sum_{m=1}^M \frac{\log V_n}{q_{X_n,m}}\delta_m(\Phi_n),\label{eq:Tn}\\
\Phi_n=& \Phi(X_n,T_n,W_n),\label{eq:Phin}
\end{align}
with $X_0=\sum_{k=1}^p k \mathbbm{1}_{[\sum_{i=1}^{k-1}\alpha_i,\sum_{i=1}^k \alpha_i )}(U_0)$ whilst the functions $\Phi$ and $F$ are defined respectively by
\begin{align*}
\Phi(x,t,w)=&\sum_{m=1}^M m \mathbbm{1}_{\big[\sum_{\ell=1}^{m-1}\phi_{x,\ell}(t), \sum_{\ell=1}^m\phi_{x,\ell}(t)\big)}(w),\\
F(x,m,u)=&\sum_{k=1}^p  k \mathbbm{1}_{\big[\sum_{w=1}^{k-1} \pi_{xw,m}, \sum_{w=1}^{k} \pi_{xw,m}\big)}(u).
\end{align*}
\end{theo}
\noindent \textbf{Proof.} It amounts to showing that the two processes (\ref{eq:mixture}) and (\ref{eq:main}) have the same conditional transition probability of moving from a current state to another. See Section \ref{sec:sec3} for details. 
$\blacksquare$ \medskip

Notice that in the absence of the regime-switching sequence $\{\Phi_n\}_{n\geq 0}$, the system dynamics (\ref{eq:main})-(\ref{eq:Tn}) coincides with that of for Markov jump process. See Ch. 5 of Resnick \cite{Resnick} for details. 

It is clear to see from their construction that although the two stochastic processes (\ref{eq:mixture}) and (\ref{eq:main}) are equivalent in distribution, they have completely different realization of their paths. In contrast to the conditional Markov jump process (\ref{eq:main}), the mixture process (\ref{eq:mixture}) never changes the regime $\Phi^{(m)}$ since it is only sampled once (at initial time $T_0$) as opposed to at every epoch time $\{T_n\}_{n\geq 0}$ for (\ref{eq:main}).

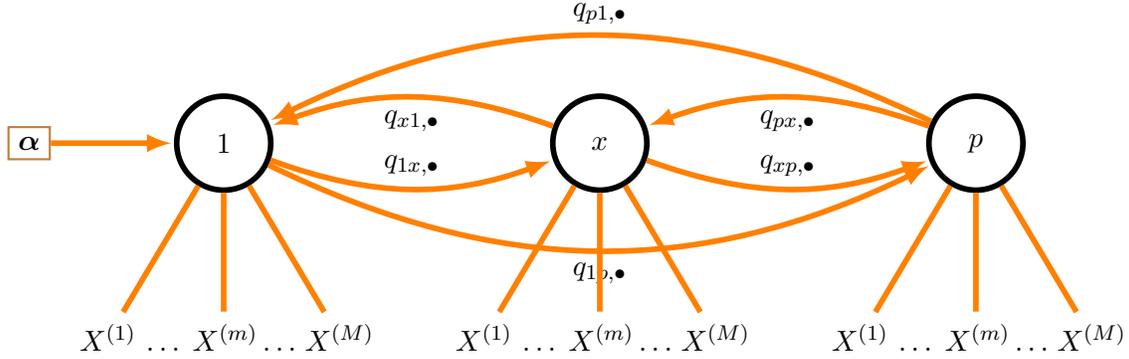
\begin{figure}[t!]
\begin{center}
  \begin{tikzpicture}[font=\sffamily]

        \tikzset{node style/.style={state,
                                    minimum width=1.25cm,
                                    line width=0.75mm,
                                    fill=white!20!white}}

          \tikzset{My Rectangle1/.style={rectangle, draw=brown, fill=white, thick,
    prefix after command= {\pgfextra{\tikzset{every label/.style={blue}}, label=below}}
    }
}

          \tikzset{My Rectangle2/.style={rectangle,draw=brown,  fill=yellow, thick,
    prefix after command= {\pgfextra{\tikzset{every label/.style={blue}}, label=below}}
    }
}

          \tikzset{My Rectangle3/.style={rectangle, draw=brown, fill=white, thick,
    prefix after command= {\pgfextra{\tikzset{every label/.style={blue}}, label=below}}
    }
}

        \node[node style] at (2, 0)     (s1)     {$1$};
        \node[node style] at (7, 0)     (s2)     {$x$};
         \node[node style] at (12, 0)    (s3)     {$p$};

      \node [My Rectangle3, label={}] at  ([shift={(-5em,0em)}]s1.west) (p0) {$\boldsymbol{\alpha}$};

        \node [draw=none, label={} ] at  ([shift={(4em,-5em)}]s1.south) (g1) {$X^{(M)}$};
         \node [draw=none, label={} ] at  ([shift={(0em,-5em)}]s1.south) (g10) {$\dots\; X^{(m)}\dots$};
          \node [draw=none, label={} ] at  ([shift={(-4em,-5em)}]s1.south)  (g2) {$X^{(1)}$};

         \node [draw=none, label={} ] at  ([shift={(4em,-5em)}]s2.south) (g3) {$X^{(M)}$};
          \node [draw=none, label={} ] at  ([shift={(0em,-5em)}]s2.south) (g30) {$\dots\; X^{(m)}\dots$};
          \node [draw=none, label={} ] at  ([shift={(-4em,-5em)}]s2.south)  (g4) {$X^{(1)}$};

         \node [draw=none, label={}, auto=right ] at  ([shift={(4em,-5em)}]s3.south) (g5) {$X^{(M)}$};
          \node [draw=none, label={} ] at  ([shift={(0em,-5em)}]s3.south) (g50) {$\dots\; X^{(m)}\dots$};
          \node [draw=none, label={} ] at  ([shift={(-4em,-5em)}]s3.south)  (g6) {$X^{(1)}$};

        \draw[every loop,
              auto=right,
              line width=0.75mm,
              >=latex,
              draw=orange,
              fill=orange]

            (s1)  edge[bend right=20, auto=left] node {$q_{1x,\bullet}$} (s2)

            (s2)  edge[bend right=20, auto=left]                    node {$q_{x1,\bullet}$}  (s1)

             (s3)  edge[bend right=20, auto=left]                    node {$q_{px,\bullet}$}  (s2)
             (s2)  edge[bend right=20, auto=left]                node {$q_{xp,\bullet}$}  (s3)

             (s3)  edge[bend right=26, auto=right]                node {$q_{p1,\bullet}$}  (s1)
             (s1)  edge[bend right=26, auto=right]                node {$q_{1p,\bullet}$}  (s3)

            (s1) edge [-,auto=left]  node {} (g1)
            (s1) edge [-,auto=left]  node {} (g10)
            (s1) edge [-,auto=right] node {} (g2)

            (s2) edge [-,auto=left] node {} (g3)
             (s2) edge [-,auto=left]  node {} (g30)
            (s2) edge [-,auto=right]  node {} (g4)

            (s3) edge [-,auto=left] node {} (g5)
             (s3) edge [-,auto=left]  node {} (g50)
            (s3) edge [-,auto=right] node {} (g6)

            (p0) edge node {} (s1);

 \end{tikzpicture}
 \caption{State diagram of the conditional MJP $X$ (\ref{eq:main}) with $M$ speed regimes.}\label{fig:mixture}
\end{center}
\end{figure}

Figure \ref{fig:mixture} exhibits a transition diagram of $X$ (\ref{eq:main}). The process selects an initial state $x_0\in\mathbb{S}=\{1,\cdots,p\}$ based on the distribution $\boldsymbol{\alpha}$ before it makes a transition from one phase of the state space $\mathbb{S}$ to another. When the process makes a transition from its current state $x$ at time $t\geq 0$ to another state, say $y$, it may switch the transition rate among $M$ possible choices $\{q_{xy,1},\cdots,q_{xy,M}\}$ with switching probability distribution $\phi_{x,\bullet}(t)=\big(\phi_{x,1}(t),\cdots,\phi_{x,M}(t)\big)$ (\ref{eq:phixt}) depending on the current state and time of the process as well as its past trajectories. The chosen rate $q_{xy,m}$ determines the underlying Markov process $X^{(m)}$ by which $X$ makes the transition. As the process is non Markovian, the transition depends on the holding time of the process in the current state. See Section \ref{sec:sec3} for details. Algorithm for simulation of sample paths of $X$ is presented in Section \ref{sec:algol}.

\begin{figure}[t!]
\centering
\subfloat[7th sample paths  of $X(t)$ (\ref{eq:main})]{\label{a1}\includegraphics[width=.5\linewidth]{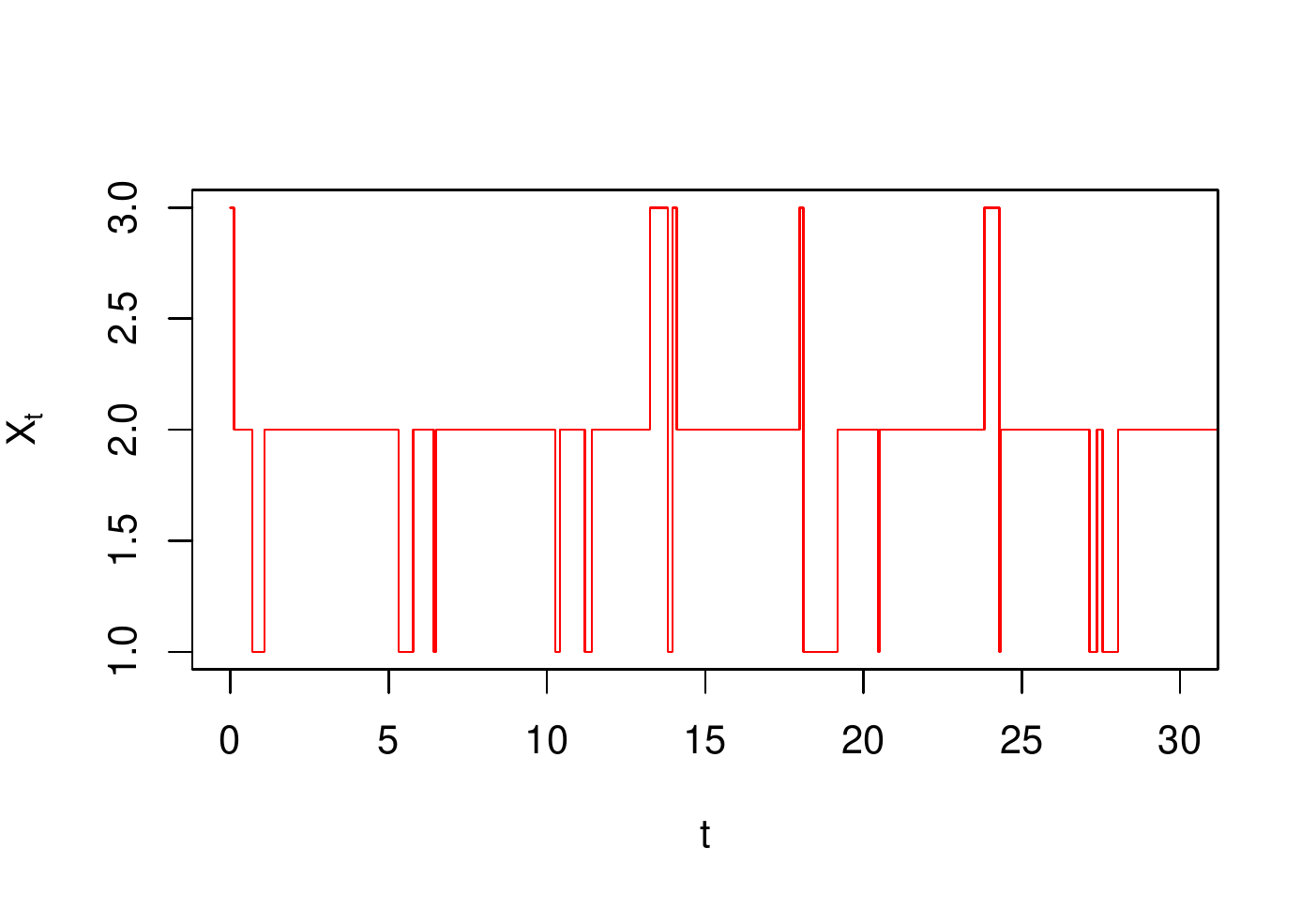}}\hfil
\subfloat[87th sample paths of $X(t)$ (\ref{eq:main})]{\label{c1}\includegraphics[width=.5\linewidth]{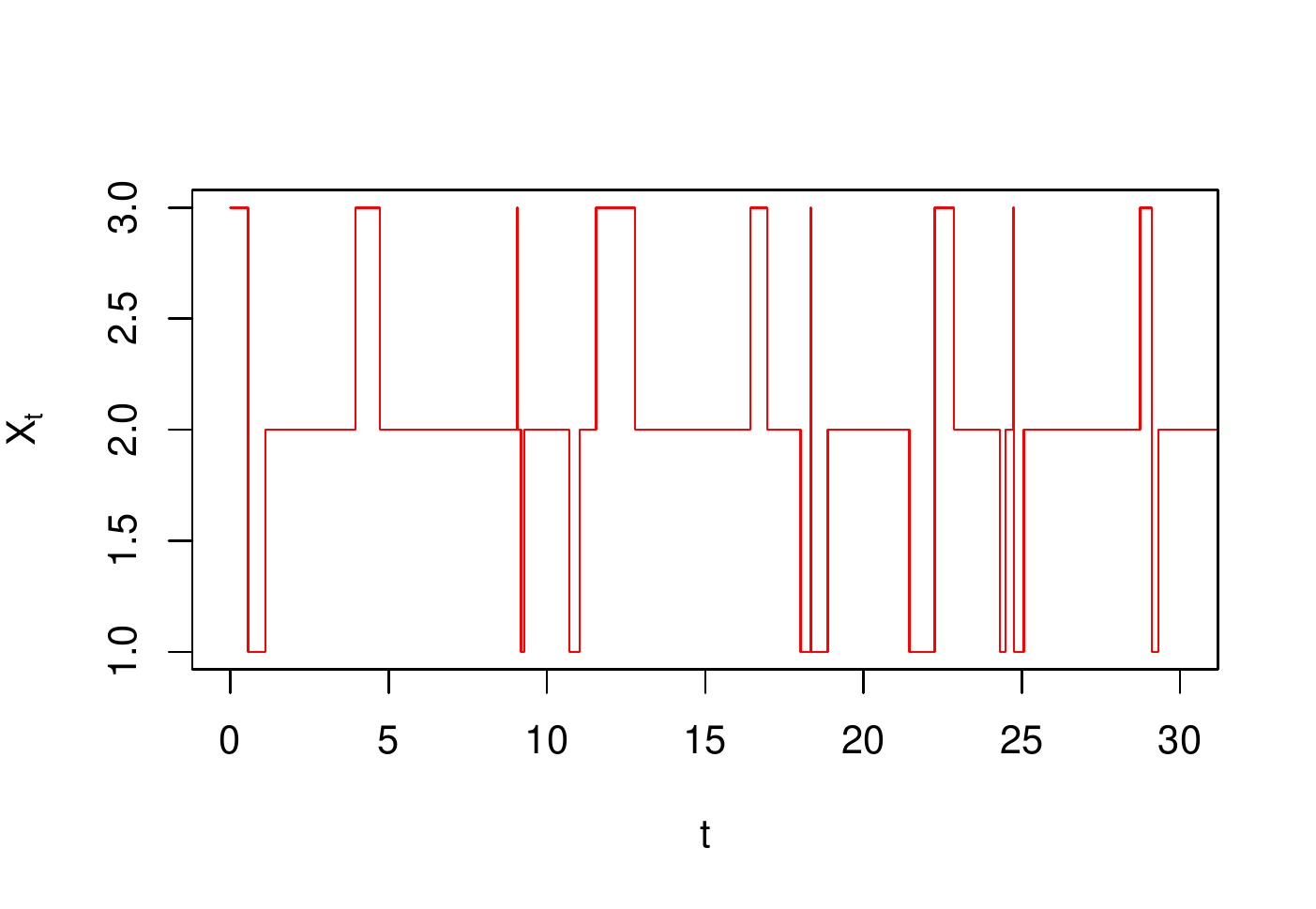}}\par
\subfloat[7th paths regime $\Phi_n$ (\ref{eq:Phin})]{\label{b1}\includegraphics[width=.5\linewidth]{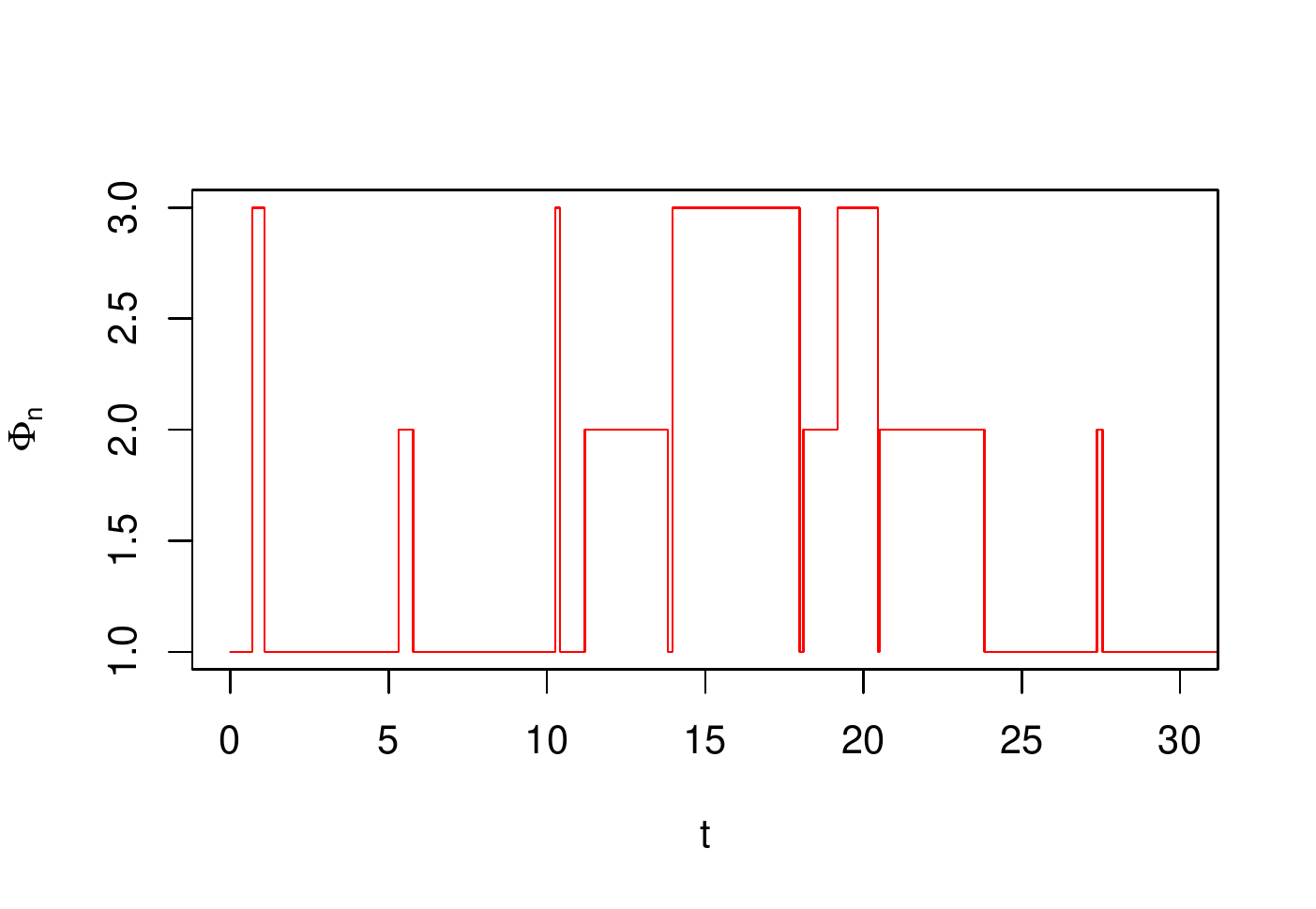}}\hfil
\subfloat[87th paths regime $\Phi_n$ (\ref{eq:Phin})]{\label{d1}\includegraphics[width=.5\linewidth]{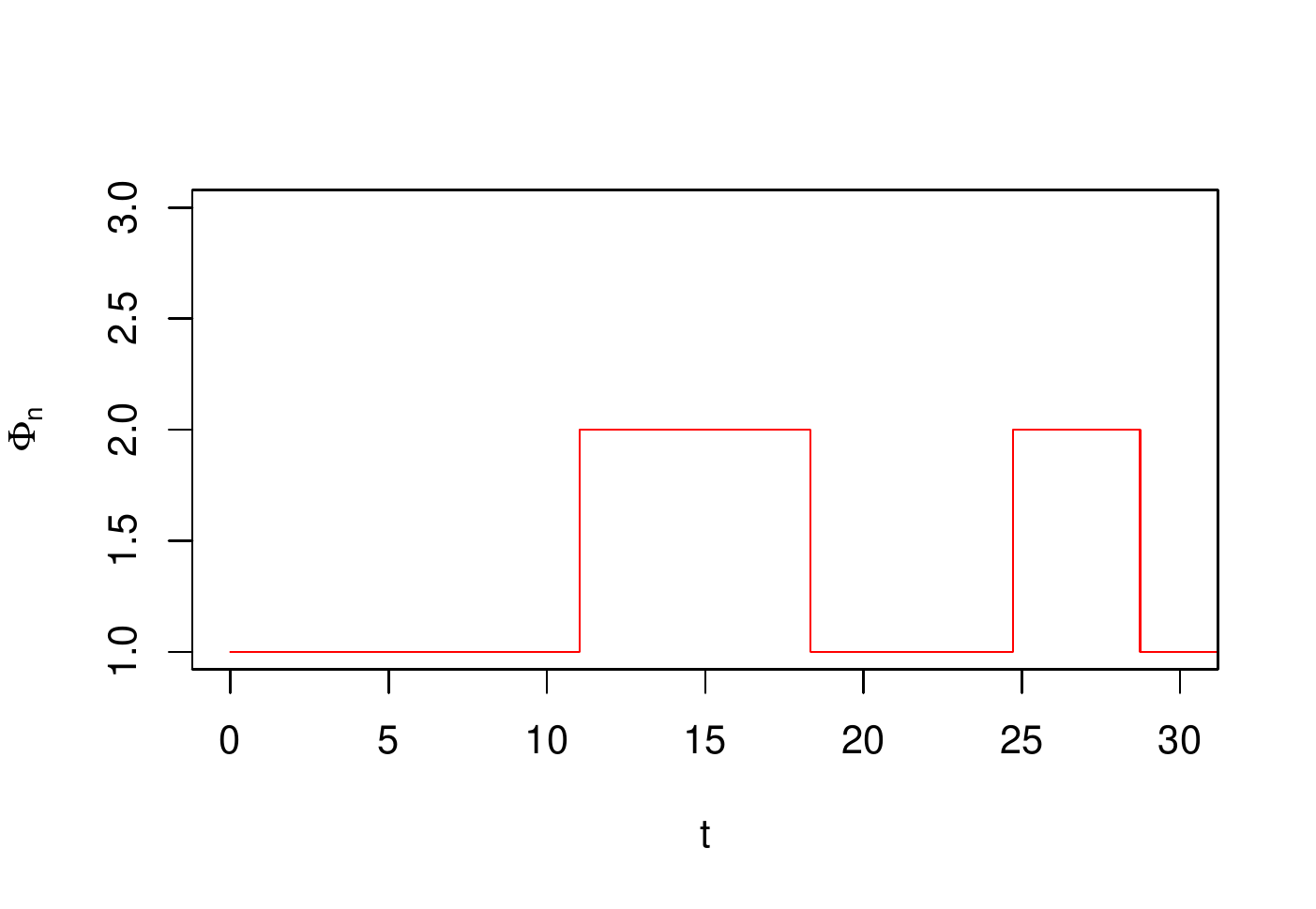}}\par
\caption{Sample paths of $X$ (\ref{eq:main}) on $[0,30]$ and the regime membership $\Phi_n$ (\ref{eq:Phin}). } \label{paths}
\end{figure}

Figure \ref{paths} displays two independent sample paths of $X$ (\ref{eq:main}) along side the respective sequence of selected speed regimes $\Phi_n$. Figure \ref{paths} (A) and (C) depicts a generated sample paths with frequent changes of speed regimes, while Figure \ref{paths} (B) and (D) are the ones with much less changes of speed regime. Note that even though the sequence $\{X_n,T_n\}$ is observable, one can not tell the sequence $\{\Phi_n\}$ of the underlying speed regimes from the observation of sample paths of $X$ on $[0,T]$.

\begin{cor}
It follows from (\ref{eq:phixt}) that, if for all $1\leq m\leq M$, $\{X^{(m)}\}$ has the same transition rate $q_{xy}>0$, i.e., $q_{xy,m}=q_{xy}$ for any $(x,y)\in\mathbb{S}$, $X$ (\ref{eq:main}) reduces to a simple Markov jump process.
\end{cor}

\section{Distributional Properties}\label{sec:sec3}

This section discusses distributional properties of the conditional Markov jump process $X$ (\ref{eq:main}) in terms of the distribution of its holding time $\tau_t$, under the measure $\mathbb{P}\{\bullet \vert X(t)=x, \mathcal{H}_{t-}\}$, defined by $$\tau_t=\inf\big\{u>t: X(u)\neq X(t)\big\},$$ the joint distribution of the holding time $\tau_t$ and the jump $X(\tau_t)$ given the current state $X(t)$ and continuous observation $\mathcal{H}_{t-}=\{X(s): 0\leq s<t\}$ of $X$ prior to $t> 0$, and the conditional transition probability of moving from one state to another. By stationary and memoryless property of the Markov jump process $X^{(m)}$, the result in Lemma \ref{lem:KBI} below shows that, conditional on the event $\{\Phi=m\}$, the transition probability $P_{xy}^{(m)}(t)$ of the Markov process $X^{(m)}$ defined by
\begin{align*}
P_{xy}^{(m)}(t)=\mathbb{P}\{X(t)= y\vert \Phi=m, X(0)=x\}, \quad \textrm{with $t\geq 0$, $x,y\in\mathbb{S}$},
\end{align*}
is the unique solution of the \textit{Kolmogorov backward integral equation}, see Proposition 5.4.1 in \cite{Resnick}. 

\begin{lem}\label{lem:KBI}
For $(x,y)\in\mathbb{S}$ and each $1\leq m\leq M$, $P_{x,y}^{(m)}(t)$ solves the integral equation
\begin{align}\label{eq:intdiff}
P_{x,y}^{(m)}(t)=e^{-q_{x,m}t}\delta_x(y) + q_{x,m}e^{-q_{x,m}t}\int_0^{t} e^{q_{x,m}u}\sum_{w\neq x} \pi_{xw,m} P_{wy}^{(m)}(u)du, \quad t\geq 0,
\end{align}
which has the unique solution given by
\begin{align*}
P_{xy}^{(m)}(t)=\mathbf{e}_x^{\top}\exp\big(Q_m t\big)\mathbf{e}_y,
\end{align*}
where $\mathbf{e}_x$ denotes a $(p\times 1)$column vector with the $x$th element equal one and zero otherwise.
\end{lem}

For the mixture process (\ref{eq:mixture}), it is straightforward to check using the Bayes formula by conditioning on the event $\{\Phi=m\}$ and Theorem 2.8.4 in Norris \cite{Norris} that the following results hold. 

\begin{prop}\label{prop:prop1}
For $u>t>0$ and $(x,y)\in\mathbb{S}$, with $y\neq x$, it holds for $M\geq 1$ that
\begin{align}
\mathbb{P}\big\{\tau_t>u \big\vert X(t)=x,\mathcal{H}_{t-}\big\}=&\sum_{m=1}^M \phi_{x,m}(t)e^{-q_{x,m}(u-t)}, \label{eq:id1}\\
\mathbb{P}\big\{\tau_t\leq u, X(\tau_t)=y \big\vert X(t)=x,\mathcal{H}_{t-}\big\}=&\sum_{m=1}^M \phi_{x,m}(t)\big[1-e^{-q_{x,m}(u-t)}\big]\frac{q_{xy,m}}{q_{x,m}}. \label{eq:id2}
\end{align}
\end{prop}

\noindent{\textbf{Proof:}} See Appendix A for the proof for the conditional Markov jump process (\ref{eq:main}). $\blacksquare$

\begin{Rem}
It is straightforward to check from (\ref{eq:id1}) and (\ref{eq:id2}) that, unless $\phi_{x,m}(t)=1$, the holding time $\tau_t$ and the jump $X(\tau_t)$ are not independent given the information set $\{X(t)=x,\mathcal{H}_{t-}\}$, i.e., $$\mathbb{P}\big\{\tau_t\leq u, X(\tau_t)=y \big\vert X(t)=x,\mathcal{H}_{t-}\big\}\neq \mathbb{P}\big\{\tau_t\leq u\big\vert X(t)=x,\mathcal{H}_{t-}\big\}\mathbb{P}\big\{X(\tau_t)=y \big\vert X(t)=x,\mathcal{H}_{t-}\big\}.$$
\end{Rem}

Interpretation of the identities (\ref{eq:id1}) and (\ref{eq:id2}) is that if $X$ (\ref{eq:main}) starts from a state $x\in\mathbb{S}$ at time $t>0$, the process would make a jump $X(\tau_t)$ independent of the length of time $\tau_t$ it stays in the state $x$ when the process selects $X^{(m)}$, with probability $\phi_{x,m}(t)$, as the underlying process to make that jump/transition from $x$ to $y=X(\tau_t)$. However, the probability of selecting the underlying process $X^{(m)}$ for making the transition dependents on the information set $\mathcal{H}_{t,x}=\{X(t)=x\}\cup\mathcal{H}_{t-}$.  

Following the two identities (\ref{eq:id1}) and (\ref{eq:id2}), one can show (see Frydman and Schuermann \cite{Frydman2005} and Surya \cite{Surya2018} for two-component mixtures, and Surya \cite{Surya2020} for general mixture) that the conditional transition probability matrix $P_{xy}(t,r)=\mathbb{P}\big\{X(r)=y \big\vert X(t)=x,\mathcal{H}_{t-}\big\}$ of $X$ (\ref{eq:mixture}) is given below. 
\begin{theo}\label{theo:transmix}
For $r\geq t\geq 0$ and $(x,y)\in\mathbb{S}$, the conditional transition probability of $X$ (\ref{eq:main}) is
\begin{align}\label{eq:transmix}
\mathbb{P}\big\{X(r)=y \big\vert X(t)=x,\mathcal{H}_{t-}\big\}=\sum_{m=1}^M \phi_{x,m}(t) P_{xy}^{(m)}(r-t).
\end{align}
\end{theo}

\noindent{\textbf{Proof:}} See Appendix B for details of the proof. $\blacksquare$

\begin{cor}\label{cor:cor1}
It follows from (\ref{eq:transmix}) that if the Markov processes $\{X^{(m)}\}_{m\geq 1}$ move at the same transition rate $q_{xy}$ for $(x,y)\in \mathbb{S}$, i.e., $q_{xy,m}=q_{xy}$ for all $1\leq m\leq M$, or equivalently $P_{xy}^{(m)}(t)=P_{xy}(t)$ with $P_{xy}(t)=\mathbf{e}_x^{\top} \exp\big(Q t) \mathbf{e}_y$, then since $\sum_{m=1}^M \phi_{x,m}(t) =1$ the identity (\ref{eq:transmix}) reduces to $$\mathbb{P}\big\{X(r)=y \big\vert X(t)=x,\mathcal{H}_{t-}\big\}=P_{xy}(r-t),$$ in which case $X$ (\ref{eq:main}) is just a simple time-homogeneous Markov jump process.
\end{cor}

\section{Maximum likelihood estimation of the distribution parameters}\label{sec:sec4}
This section discusses maximum likelihood estimation of distribution parameters of the conditional Markov jump process $X$ (\ref{eq:main}) based on continuous observation of the sample paths. To be more precise, we are interested in estimating the initial regime switching probability $\phi_{x,m}$ and the intensity matrices $Q_m=[q_{xy,m}]_{xy}$, for $(x,y)\in \mathbb{S}$ and $1\leq m\leq M$ from continuously observed realizations of the process. In general, the estimation problem is challenging and quite difficult given that the process changes its speed/underlying process at random times. However, thanks to their distributional equivalence with the mixture of Markov jump processes (\ref{eq:mixture}), the estimation can be done systematically using the EM algorithm with an explicit form of parameter estimates. Estimated variance of the parameter estimates are discussed along with their asymptotic properties. 

\subsection{Likelihood function of continuously observed sample paths}
To write the likelihood function of the observed sample paths, let $X^{k}=\{X^k(t), 0\leq t\leq T<\infty\}$ denote the k'th realization of $X$ (\ref{eq:main}) on $[0,T]$ with $T$ being the end of observation time.  To write the likelihood of $X^k$, for $1\leq k\leq K$ and $1\leq m\leq M$, we consider the following statistics associated with the paths $X^k$:
\begin{equation}\label{eq:stats}
\begin{split}
\Phi _{k,m} =&I(X^k=X^{(m)}) \quad \textrm{with \; $\sum_{m=1}^M \Phi_{k,m}=1$},   \\
B_{x}^{k} =&I(X_{0}^{k}=x), \; x\in E,  \\
N_{xy}^{k} =&\#\text{ of times }X^{k}\text{ makes an }x\rightarrow y \text{
transition, }y\neq x, x\in E,  \\
T_{x}^{k} =&\int_{0}^{T}I(X_{u}^{k}=x)du=\text{ total time }X^{k}\text{
spends in state }x\in E,  
\end{split}
\end{equation}
where $\Phi _{k,m}$ is equal to 1 if $X^k$ evolves according to the Markov process $X^{(m)}$ and equal to zero, otherwise. We denote by $\Phi^k$ the corresponding (categorical) random variable for $X^k$ s.t. $\{\Phi^k=m\}=\{\Phi_{k,m}=1\}$. Note that both $\Phi_{k,m}$ and $\Phi^k$ are unknown, but other quantities are known. 

\begin{Rem}
For estimation, the above statistics need to be computed for each sample paths $X^k$. Unlike in the case of Markov process, they do not represent an aggregate for all paths $\{X^k\}_{k\geq 1}$.
\end{Rem}

Suppose that $X^k$ chooses its initial state $x_0\in\mathbb{S}$ to start with randomly at probability $\alpha_{x_0}$. On account that the bivariate process $(X^k,\Phi^k)$ is Markovian, it follows from the likelihood (\ref{eq:likelihood})
\begin{align*}
f_{\theta}(X^k,\Phi^{k}=m)=&f_{\theta}(X^k(0)=x_0)f_{\theta}\big(\Phi^{k}=m \vert X^k(0)=x_0\big) f_{\theta}\big(X^k \vert \Phi^{k}=m, X^k(0)=x_0\big)  \nonumber\\[6pt]
=& \prod_{x=1}^p\big( \alpha_{x} \phi_{x,m}\big)^{B_{x}^{k}} \prod_{x=1}^p \prod_{y\neq x,y=1}^p \big(q_{xy,m}\big)^{N_{xy}^k} e^{-q_{xy,m} T_x^k},
\end{align*}
with $p=\vert E \vert$ and $\theta=\big(\phi_{x,m},q_{xy,m}:(x,y)\in\mathbb{S}, 1\leq m\leq M\big)$ representing the distribution parameter of the process. Hence, the likelihood function of the complete information $(X^k,\Phi^k)$ is given by
\begin{align*}
f_{\theta}(X^k,\Phi^k)= \prod_{m=1}^M \big[f_{\theta}(X^k,\Phi^k=m)\big]^{\Phi_{k,m}}.
\end{align*}

The log-likelihood of all realized independent sample paths $\mathcal{X}=\cup_{k=1}^K X^k$ is represented by
\begin{align}\label{eq:likelihood3}
\log f_{\theta}(\mathcal{X})=&\log\prod_{k=1}^K\sum_{m=1}^M f_{\theta}(X^k,\Phi^k=m) \nonumber\\
=&\sum_{x=1}^p \sum_{k=1}^K B_x^k \log \alpha_{x} - K\Big(\sum_{x=1}^p \alpha_x-1\Big) \\
&\hspace{1cm}+ \sum_{k=1}^K \log \Big(\sum_{m=1}^M \Big[\prod_{x=1}^p \phi_{x,m}^{B_{x}^{k}} \prod_{x=1}^p \prod_{y\neq x,y=1}^p \big(q_{xy,m}\big)^{N_{xy}^k} e^{-q_{xy,m} T_x^k}\Big]\Big), \nonumber
\end{align}
where we have taken into account the constraint $\sum_{x=1}^p \alpha_x=1$ in the log-likelihood function. It follows from the log-likelihood $\log f_{\theta}(\mathcal{X})$ (\ref{eq:likelihood3}) that the estimate $\widehat{\alpha}_x^0$ for the true initial probability $\alpha_x^0$ can be easily obtained separately from the other parameters so that if $K$ realizations are available,
\begin{align*}
\widehat{\alpha}_x^0=\frac{1}{K}\sum_{k=1}^K B_x^k.
\end{align*}
Therefore, it is excluded from the estimation of other parameters discussed in the section below. 

\subsection{Maximum likelihood estimation}

Since neither information on $\Phi^k$ nor $\Phi_{k,m}$ is available from the observed sample paths $X^k$, we can not directly use the likelihood (\ref{eq:likelihood3}) to estimate the parameters $\phi_{x,m}$ and $q_{xy,m}$ for $(x,y)\in\mathbb{S}$ and $1\leq m\leq M$. Moreover, maximizing the log-likelihood $\log f_{\theta}(\mathcal{X})$ (\ref{eq:likelihood3}) does not lead to an explicit estimate for $\phi_{x,m}$ and $q_{xy,m}$. Instead, we consider the sample average of the conditional expectation of $\log f_{\theta}(X^k,\Phi^k)$ given only the sample paths $X^k$:
\begin{eqnarray*}
L_K(\theta)=\frac{1}{K}\sum_{k=1}^K E_{\theta_0}\Big(\log f_{\theta}(X^k,\Phi^k)\Big\vert X^k\Big), 
\end{eqnarray*}
where $\mathbb{E}_{\theta_0}$ refers to the expectation operator associated with the probability measure $\mathbb{P}_{\theta_0}$ under which the paths $\{X^k\}$ were generated under the true parameter value $\theta_0=(\phi^0,q^0)$ of $\theta=(\phi,q)$. 

Given the constraint $\sum_{m=1}^M \phi_{x,m}=1$, for each $x\in\mathbb{S}$, the log-likelihood of $(X^k,\Phi^k)$ reads as
\begin{align}\label{eq:loglike}
\log f_{\theta}(X^k,\Phi^k)=&\sum_{m=1}^M\sum_{x=1}^p \Phi_{k,m}B_x^k \log \phi_{x,m}-\sum_{x=1}^p B_x^k\Big(\sum_{m=1}^M \phi_{x,m}-1\Big) \nonumber\\
&+\sum_{m=1}^M\sum_{x=1}^p\Big(\sum_{y\neq x, y=1}^p \Phi_{k,m}N_{xy}^k \log q_{xy,m} -\Big(\sum_{y\neq x,y=1}^p q_{xy,m}\Big)\Phi_{k,m}T_x^k\Big).
\end{align}
For the maximum likelihood estimation of $\theta_0$ discussed below, define $\widehat{\Phi}_{k,m}(\theta)=\mathbb{E}_{\theta}\big(\Phi_{k,m}\big\vert X^k\big)$, i.e.,
\begin{align*}
\widehat{\Phi}_{k,m}(\theta)=\frac{f_{\theta}(X^k,\Phi^k=m)}{\sum_{m=1}^M f_{\theta}(X^k,\Phi^k=m)}.
\end{align*}

One can verify using (\ref{eq:loglike}) that the following regularity conditions hold for each element $\theta_i$ of $\theta$,
\begin{align}\label{eq:bounded}
\frac{1}{K}\sum_{k=1}^K\mathbb{E}_{\theta_0}\Big(\Big\vert \frac{\partial}{\partial \theta_i}\log f_{\theta}(X_k,\Phi_k)     \Big\vert \Big\vert X_k\Big)<\infty, \quad \textrm{for $K\geq 1$}.
\end{align}
To be more precise, for $\phi_{x,m}$ the upper bound is $1+1/\phi_{x,m}$, and is $\frac{1}{K}\sum_{k=1}^K \big( N_{xy}^k/q_{xy,m} + T_x^k\big)$ for $q_{xy,m}$. Let $\Theta$ be the set of any possible value, other than zero, of $\theta_i$. By the Bayes' formula, it is not hard to show that the following result holds for all $\theta_i$. The result is needed to derive the MLE of the distribution parameters and the observed Fisher information matrix of the MLE.
\begin{prop}\label{prop:mainprop}
For a given sample paths of $X$, it holds true for any $\theta_i\in\Theta$ that
\begin{align}\label{eq:id10}
\frac{\partial }{\partial \theta_i} \log f_{\theta}(X)=\mathbb{E}_{\theta}\Big( \frac{\partial }{\partial \theta_i}\log f_{\theta}(X,\Phi)\Big\vert X  \Big).
\end{align}
\end{prop}
\noindent \textbf{Proof.}
By Bayes' formula for conditional probability, the conditional expectation reads as
\begin{align*}
\mathbb{E}_{\theta}\Big(\frac{\partial}{\partial \theta_i} \log f_{\theta}(X,\Phi) \Big\vert X \Big)=&\sum_{m=1}^{M} \frac{\partial}{\partial \theta_i} \log f_{\theta}(X,\Phi=m)f_{\theta}(\Phi=m\big\vert X)\\
=&\frac{1}{f_{\theta}(X)}\sum_{m=1}^M \frac{\partial}{\partial \theta_i} f_{\theta}(X,\Phi=m),
\end{align*}
from which we arrive at the claim on account that $f_{\theta}(X)=\sum_{m=1}^M f_{\theta}(X,\Phi=m)$. $\blacksquare$

\begin{theo}
The estimate $\widehat{\theta}_0$ solving the equation $L_K^{\prime}(\theta_0)=0$ coincides with the MLE, i.e., $$\frac{\partial}{\partial \theta}\log f_{\theta}(\mathcal{X})=\sum_{k=1}^K \frac{\partial}{\partial \theta} \log f_{\theta}(X^k)=0 \quad \textrm{at $\theta=\widehat{\theta}_0$.}$$
Furthermore, the MLE of the true unknown parameters $\phi_{x,m}^0$ and $q_{xy,m}^0$ are respectively
\begin{align}\label{eq:theMLE}
\widehat{\phi}_{x,m}^0=\frac{\sum_{k=1}^K \widehat{\Phi}_{k,m}(\widehat{\theta}_0) B_x^k}{\sum_{k=1}^K B_x^k} \quad \textrm{and} \quad \widehat{q}_{xy,m}^0=\frac{\sum_{k=1}^K \widehat{\Phi}_{k,m}(\widehat{\theta}_0) N_{xy}^k}{\sum_{k=1}^K \widehat{\Phi}_{k,m}(\widehat{\theta}_0) T_{x}^k}.
\end{align}
%
\end{theo}
\noindent \textbf{Proof.}
By the regularity condition (\ref{eq:bounded}) and the result of Proposition \ref{prop:mainprop}, we have
\begin{align*}
0=L_K^{\prime}(\theta_0)=\frac{1}{K}\sum_{k=1}^K \mathbb{E}_{\theta_0}\Big(\frac{\partial}{\partial \theta} \log f_{\theta}(X^k,\Phi^k)\Big\vert X^k\Big)\Big\vert_{\theta=\theta_0}=\frac{1}{K}\sum_{k=1}^K \frac{\partial}{\partial \theta} \log f_{\theta}(X^k)\Big\vert_{\theta=\theta_0},
\end{align*}
which establishes the first part of the claims. By differentiating (\ref{eq:loglike}) w.r.t $\phi_{x,m}$ and $q_{xy,m}$, one has
\begin{align*}
\frac{\partial}{\partial \phi_{x,m}} \log f_{\theta}(X^k,\Phi^k)=&\frac{\Phi_{k,m}B_x^k}{\phi_{x,m}}-B_x^k, \\
\frac{\partial}{\partial q_{xy,m}} \log f_{\theta}(X^k,\Phi^k)=&\frac{\Phi_{k,m}N_{xy}^k}{q_{xy,m}}-\Phi_{k,m}T_x^k,
\end{align*}
respectively. On account that $B_x^k, N_{xy}^k,$  and $T_x^k$ are observable from the sample paths $X^k$, the MLE $\widehat{\phi}_{x,m}^0$ and $\widehat{q}_{xy,m}^0$ are found by solving for each parameter the equation $L_K^{\prime}(\theta_0)=0$. $\blacksquare$

\subsection{EM algorithm for the estimation}

Recall that the MLE $\widehat{\theta}_0$ as solution of $L_K^{\prime}(\theta_0)=0$ is nested in terms of $\widehat{\Phi}_{k,m}(\widehat{\theta}_0)$. Thus, it shall be found iteratively using the EM algorithm introduced by Dempster et al. \cite{Dempster}. The first step of each iteration, the \textbf{E-step}, consists of calculating the expectation $\widehat{\Phi}_{k,m}(\theta^{\ell})=\mathbb{E}_{\theta^{\ell}}\big(\Phi_{k,m}\big\vert X^k\big)$ based on the current estimate $\theta^{\ell}$ to get $L_K(\theta)$. Then in the second step, the \textbf{M-step}, the function $L_K(\theta)$ is maximized to obtain $\theta^{\ell +1}$. The two steps result in  
\begin{align}\label{eq:EMalgol}
\phi_{x,m}^{\ell+1}=\frac{\sum_{k=1}^K \widehat{\Phi}_{k,m}(\theta^{\ell}) B_x^k}{\sum_{k=1}^K B_x^k} \quad \textrm{and} \quad q_{xy,m}^{\ell+1}=\frac{\sum_{k=1}^K \widehat{\Phi}_{k,m}(\theta^{\ell}) N_{xy}^k}{\sum_{k=1}^K \widehat{\Phi}_{k,m}(\theta^{\ell}) T_{x}^k}.
\end{align}

In this way, for any $\theta\in \Theta$, $L_K(\theta^{\ell+1}) \geq L_K(\theta)$ so that the log-likelihood function $\log f_{\theta}(\mathcal{X}))$ is increasing in each iteration, i.e., $\log f_{\theta^{\ell +1}}(\mathcal{X})\geq f_{\theta^{\ell}}(\mathcal{X})$. To be more precise, by the Bayes formula, one would obtain $\log f_{\theta}(X^k)=\mathbb{E}_{\theta^{\ell}}\big(\log f_{\theta}(X^k,\Phi^k)\big\vert X^k\big)- 
\mathbb{E}_{\theta^{\ell}}\big(\log f_{\theta}(X^k,\Phi^k\big\vert X^k)\big\vert X^k\big)$. Hence,
\begin{align*}
\frac{1}{K}\sum_{k=1}^K \log f_{\theta}(X^k)=L_K(\theta)- \frac{1}{K}\sum_{k=1}^K\mathbb{E}_{\theta^{\ell}}\Big(\log f_{\theta}(X^k,\Phi^k\big\vert X^k)\Big\vert X^k\Big).
\end{align*}
By the concavity of the logarithmic function, an application of Jensen's inequality results in
\begin{align*}
&\mathbb{E}_{\theta^{\ell}}\Big(\log f_{\theta^{\ell +1}}(X^k,\Phi^k\big\vert X^k)\Big\vert X^k\Big) - \mathbb{E}_{\theta^{\ell}}\Big(\log f_{\theta^{\ell}}(X^k,\Phi^k\big\vert X^k)\Big\vert X^k\Big)\\
&\hspace{1cm}=\mathbb{E}_{\theta^{\ell}}\Big(  \log\Big[ \frac{ f_{\theta^{\ell +1}}(X^k,\Phi^k\big\vert X^k)}{ f_{\theta^{\ell}}(X^k,\Phi^k\big\vert X^k)}\Big]\Big\vert X^k\Big)\leq \log \mathbb{E}_{\theta^{\ell}}\Big(  \frac{ f_{\theta^{\ell +1}}(X^k,\Phi^k\big\vert X^k)}{ f_{\theta^{\ell}}(X^k,\Phi^k\big\vert X^k)}\Big\vert X^k\Big)=0,
\end{align*}
which implies that the log-likelihood $\log f_{\theta}(\mathcal{X})$ (\ref{eq:likelihood3}) increases in each iteration. The iteration stops when a stopping criterion $\lvert\lvert \theta^{\ell+1}-\theta^{\ell}\rvert\rvert<\varepsilon$ is satisfied, under a Euclidean norm $\lvert\lvert \bullet \rvert\rvert$. Furthermore, it follows from (\ref{eq:loglike}) that $\mathbb{E}_{\theta_0}\big(\log f_{\theta}(X^k,\Phi^k)\big\vert X^k\big)$ is continuous in both $\theta_0$ and $\theta$ away from zero. In the M-step of each iteration, $\theta$ is chosen as such that the function $\mathbb{E}_{\theta_0}\big(\log f_{\theta}(X^k,\Phi^k)\big\vert X^k\big)$ is maximized. The latter and the continuity of $\mathbb{E}_{\theta_0}\big(\log f_{\theta}(X^k,\Phi^k)\big\vert X^k\big)$ in both $\theta_0$ and $\theta$ are required for the convergence of the EM algorithm (\ref{eq:EMalgol}) to a local maximum $\widehat{\theta}_0$. See Theorem 3 in Wu \cite{Wu}.

Suppose that we have $K$ independent observations $\{X^k\}$ of the process $X$ (\ref{eq:main}), either generated sample paths or real dataset. The EM algorithm for the parameter estimation is detailed below.

\medskip

\begin{enumerate}

\item[(i)][\textbf{Initial step}] Set the number of regime $M\geq 1$. Use the same initialization of the EM algorithm \cite{Frydman2020} by randomly splitting the observations $\{X^k\}$ into $M$ subset of observations with the size $\lfloor K/M\rfloor$ for the first $M-1$ subsets and $K-(M-1)\lfloor K/M\rfloor$ for the $M$th subset. Assuming each subset comprises sample paths of time-homogeneous Markov jump processes, use each subset to estimate the intensity matrix $Q_m^0$. Use the estimate $\widehat{Q}_m^0$ for initial value $Q_m^0$. The estimate $\widehat{\alpha}_x^0$ is used for the initial distribution $\alpha_x^0$. In each state $x\in\mathbb{S}$, the initial value of $(\phi_{x,m}^0: 1\leq m\leq M)$ is set to be uniform, i.e., $\phi_{x,\bullet}^0=(1/M,\cdots,1/M)$. 

\medskip

\item[(ii)] \textbf{E-step}, after $\ell$th iteration, evaluate the conditional expectation $\widehat{\Phi}_{k,m}(\theta^{\ell})=\mathbb{E}_{\theta^{\ell}}\big(\Phi_{k,m}\big\vert X^k\big)$.

\medskip

\item[(iii)] \textbf{M-step}, get an update $\theta^{\ell +1}$ using the identity (\ref{eq:EMalgol}) and the current estimate $\theta^{\ell}$.

\medskip

\item[(iv)] Stop if $\vert\vert \theta^{\ell +1} - \theta^{\ell} \vert\vert <\varepsilon$, with $\varepsilon>0$. Otherwise, go back to (ii) and replace $\theta^{\ell}$ by $\theta^{\ell+1}.$

\end{enumerate}

\medskip

Repeat the EM algorithm for different number of regime membership $M\geq 1$ until the selected model has the lowest Akaike's information criterion (AIC)  \cite{Akaike}. See Section \ref{sec:sec6} for detailed examples.

\subsection{Observed Fisher information matrix}
Based on observation of the sample paths $\mathcal{X}$, the observed information matrix $J_p(\theta):=-\frac{\partial^2}{\partial \theta^2}\log f_{\theta}(\mathcal{X})$ can be derived using the identity (\ref{eq:id10}).
\begin{theo}\label{theo:Fisher}
For any $\theta_i,\theta_j\in \Theta$, the $(i,j)$ component of information matrix $J_p(\theta)$ is given by
\begin{align}\label{eq:InfMatr}
J_p(\theta_i,\theta_j):=&-\frac{\partial^2}{\partial \theta_i \partial \theta_j}\log f_{\theta}(\mathcal{X}) \nonumber\\
=&-\sum_{k=1}^K \mathbb{E}_{\theta}\Big(\frac{\partial^2}{\partial \theta_i \partial \theta_j} \log f_{\theta}(X^k,\Phi^k)\Big\vert X^k\Big) \nonumber\\
&\hspace{0cm}-\sum_{k=1}^K \mathbb{E}_{\theta}\Big( \Big[\frac{\partial}{\partial \theta_i}\log f_{\theta}(X^k,\Phi^k)\Big]
\Big[\frac{\partial}{\partial \theta_j}\log f_{\theta}(X^k,\Phi^k)\Big]\Big\vert X^k\Big)\\
&\hspace{0cm}+\sum_{k=1}^K\mathbb{E}_{\theta}\Big( \frac{\partial}{\partial \theta_i}\log f_{\theta}(X^k,\Phi^k)\Big\vert X^k\Big)\mathbb{E}_{\theta}\Big( \frac{\partial}{\partial \theta_j}\log f_{\theta}(X^k,\Phi^k)\Big\vert X^k\Big). \nonumber
\end{align}
\end{theo}

\noindent \textbf{Proof:} See Appendix \ref{sec:Fisher} for details of the derivation. $\blacksquare$

\medskip

The information matrix $J_p(\theta)$ (\ref{eq:InfMatr}) agrees with the general matrix formula of Louis \cite{Louis} after adding and subtracting $\Big[\sum_{k=1}^K \mathbb{E}_{\theta}\Big(\frac{\partial}{\partial \theta_i}\log f_{\theta}(X^k,\Phi^k)\Big\vert X^k\Big)\Big]\Big[\sum_{\ell=1}^K \mathbb{E}_{\theta}\Big(\frac{\partial}{\partial \theta_j}\log f_{\theta}(X^{\ell},\Phi^{\ell})\Big\vert X^{\ell}\Big)\Big]$ to (\ref{eq:InfMatr}). Therefore, (\ref{eq:InfMatr}) is the simplified form of the Louis' matrix formula, see eqn. (\ref{eq:InfMatr2}) of Appendix \ref{sec:subsecC}. When compared to the formula (\ref{eq:InfMatr2}), the convergence of $J_p(\theta_i,\theta_j)$ to the respective expected information matrix $I_p(\theta_i,\theta_j)=\mathbb{E}_{\theta}\Big(-\frac{\partial^2}{\partial \theta_i \partial \theta_j}\log f_{\theta}(X)\Big)$ is more immediate from (\ref{eq:InfMatr}) than (\ref{eq:InfMatr2}).

\begin{cor}
By identity (\ref{eq:id10}), $-\frac{1}{K}\sum_{k=1}^K \frac{\partial^2}{\partial \theta_i \partial \theta_j}\log f_{\theta}(X^k)\stackrel{\mathbb{P}}{\Longrightarrow} \mathbb{E}_{\theta}\Big(-\frac{\partial^2}{\partial \theta_i \partial \theta_j}\log f_{\theta}(X)\Big)$.
\end{cor}

From the log-likelihood function (\ref{eq:loglike}) and (\ref{eq:InfMatr}), the $(i,j)-$elements $J_p(\widehat{\theta}_i,\widehat{\theta}_j)$ of the information matrix $J_p(\widehat{\theta})$ is given below. The results coincide with Proposition 2 in Frydman and Surya \cite{Frydman2020}.

\begin{prop}\label{prop:fisher}
\label{prop:prop3} Let $A_{xy,m}^k=N_{xy}^k-\widehat{q}_{xy,m}T_x^k$. If $\widehat{\phi}_{x,m}, \widehat{q}_{xy,n}\neq 0$, for $(x,y)\in S$, $1\leq n,m\leq M$, 
\begin{eqnarray*}
J_p(\widehat{\phi}_{y,n},\widehat{\phi}_{x,m})=\delta_x(y)\sum_{k=1}^K \left(\frac{\widehat{\Phi}_{k,n}(\widehat{\theta})}{\widehat{\phi}_{y,n}} \right) %
\left(\frac{\widehat{\Phi}_{k,m}(\widehat{\theta})}{\widehat{\phi}_{x,m}}\right) B_y^k, 
\end{eqnarray*}
\begin{eqnarray*}
J_p(\widehat{q}_{x^{\prime}y^{\prime},n},\widehat{q}_{xy,m})=
-\frac{1}{\widehat{q}_{x^{\prime }y^{\prime },n}\widehat{q}_{xy,m}}%
\sum_{k=1}^{K}\widehat{\Phi }_{k,n}(\widehat{\theta})\Big(\delta_m(n)-\widehat{\Phi }_{k,m}(\widehat{\theta})\Big)A_{x^{\prime }y^{\prime },n}^{k}A_{xy,m}^{k}, 
\end{eqnarray*}
in particular, the diagonal element $J_p(\widehat{q}_{xy,m}):=J_p(\widehat{q}%
_{xy,m},\widehat{q}_{xy,m})$ is given by 
\begin{eqnarray*}
J_p(\widehat{q}_{xy,m})=\frac{\widehat{N}_{xy,m}}{\widehat{q}_{xy,m}^{2}}-
\frac{1}{\widehat{q}_{xy,m}^{2}}\sum_{k=1}^{K}\widehat{\Phi }_{k,m}(\widehat{\theta})\big(1-
\widehat{\Phi }_{k,m}(\widehat{\theta})\big)\Big( A_{xy,m}^{k}\Big) ^{2},
\end{eqnarray*}
\begin{eqnarray*}
J_p(\widehat{\phi }_{x,m},\widehat{q}_{x^{\prime }y,n})=
-\frac{1}{\widehat{\phi }_{x,m}\widehat{q}_{x^{\prime }y,n}}\sum_{k=1}^{K}%
\widehat{\Phi }_{k,n}(\widehat{\theta})\Big(\delta_m(n)-\widehat{\Phi }_{k,m}(\widehat{\theta})\Big)A_{x^{\prime
}y,n}^{k}B_{x}^{k}.
\end{eqnarray*}
\end{prop}

\noindent \textbf{Proof:} See Appendix D for details of the derivation. $\blacksquare$

\medskip

The information matrix $J_p(\widehat{\theta})$ is used to obtain an estimate of the standard errors of the MLE $\widehat{\theta}$. To be more precise, the estimated standard error $\widehat{\sigma}_{\widehat{\theta}_i}:=\sqrt{\widehat{\textrm{Var}}(\widehat{\theta}_i)}$ corresponds to the square root of $i$th diagonal element of the inverse $J_p^{-1}(\widehat{\theta})$, i.e., $\widehat{\sigma}_{\widehat{\theta}_i}=\sqrt{\big[J_p^{-1}(\widehat{\theta})\big]_{ii}}$. See Section \ref{sec:sec6} for applications.

\begin{Rem}
To prevent having singularity in taking the inverse of information matrix $J_p(\widehat{\theta})$, we exclude from $\widehat{\theta}$ a (very close to) zero estimate $\widehat{\theta}_i$ of $\theta_i$, for some $i=1,\cdots,n$, and replace $\theta_i$ by 0.
\end{Rem}

In the section below, we derive the large sample properties of the MLE $\widehat{\theta}_0$ (\ref{eq:theMLE}).

\section{Large sample properties of the MLE $\widehat{\theta}_0$}\label{sec:sec5}

\subsection{\textbf{Consistency and asymptotic normality of the MLE $\widehat{\theta}_0$}}

On account of Lemma B2 and Proposition C1 in Frydman and Surya \cite{Frydman2020} and Theorem 1 in Van Loan \cite{VanLoan}, which gives an exact solution to a (double) definite integral involving exponential of intensity matrices, one can show for all $(x,y)\in\mathbb{S}$ and $1\leq m\leq M$ that the first two unconditional moments of $\Phi_{k,m}B_x^k$, $\Phi_{k,m}N_{xy}^k$, and $\Phi_{k,m}T_x^k$ are finite for every $1\leq k\leq K$. These would imply that $\mathbb{E}_{\theta}\big(\log f_{\theta}(X^k,\Phi^k)\big)<\infty$,
\begin{align}\label{eq:regular}
\mathbb{E}_{\theta}\Big( \frac{\partial}{\partial \theta_i} \log f_{\theta}(X^k,\Phi^k)\Big)<\infty, \quad \textrm{and} \quad \mathbb{E}_{\theta}\Big( \frac{\partial^2}{\partial \theta_i \partial \theta_j} \log f_{\theta}(X^k,\Phi^k)\Big)<\infty.
\end{align}

The following results present explicit form of the expected Fisher information matrix $$I_c(\theta_i,\theta_j)=-\mathbb{E}_{\theta}\Big(\frac{\partial^2}{\partial \theta_i \partial \theta_j} \log f_{\theta}(X,\Phi)\Big),$$ under complete observation $(X,\Phi)$. The result shows that the matrix $I_c(\theta)$ is of block diagonal matrix form. It is used in the asymptotic distribution and asymptotic efficiency of the MLE $\widehat{\theta}$.

\begin{prop}\label{prop:MatIc}
For $(x,x^{\prime},y,y^{\prime})\in\mathbb{S}$ and $1\leq n,m\leq M$, with $\alpha_x^0,\phi_{x,m}^0, q_{x^{\prime}y^{\prime},n}^0\neq 0$, $I_c(\theta_0)$ is
\begin{align}
I_c(\phi_{x,m}^0,\phi_{y,n}^0)=&\frac{\alpha_y^0}{\phi_{x,m}^0}\delta_x(y)\big(\delta_m(n)-\phi_{x,m}^0\big),  \nonumber\\
I_c(\phi_{x^{\prime},n}^0,q_{xy,m}^0)=&0,  \label{eq:expfisher}\\
I_c(q_{xy,m}^0,q_{x^{\prime}y^{\prime},n}^0)=& \frac{\delta_m(n)\delta_x(x^{\prime})\delta_y(y^{\prime})}{q_{x^{\prime}y^{\prime},n}^0} \alpha_0^{\top} S_n^0 \int_0^T e^{Q_n^0 u }e_x du, \nonumber
\end{align}
where $S_n^0=\textrm{diag}\big(\phi_{1,n}^0,\cdots,\phi_{p,n}^0\big)$, $Q_n^0=[q_{xy,n}^0]_{xy}$, and $\alpha_0=(\alpha_1^0,\cdots,\alpha_p^0)$.
\end{prop}
\noindent \textbf{Proof:} See Appendix E for details of the derivation. $\blacksquare$

\medskip

By independence of the sample paths $\{X^k\}$ it follows by applying the law of large numbers that the function $L_K(\theta)$ converges with probability one as $K\rightarrow \infty$ to $$L(\theta)=\mathbb{E}_{\theta_0}\big(\log f_{\theta}(X,\Phi)\big),$$
where $(X, \Phi)$ is a generic representations of $(X^k,\Phi^k)$. The result below, which is based on Proposition 4 in Frydman and Surya \cite{Frydman2020}, shows that the function $L(\theta)$ has a global maximum at $\theta=\theta_0$.
\begin{prop}\label{prop:LT}
For any $\theta\in\Theta$, $L(\theta)\leq L(\theta_0)$. Moreover, $L(\theta)<L(\theta_0)$, unless $\mathbb{P}_{\theta_0}\big\{f_{\theta}(X,\Phi)=f_{\theta_0}(X,\Phi)\big\}=1$. Furthermore, the equation $L^{\prime}(\theta)=0$ has the unique solution $\theta=\theta_0$.
\end{prop}

The results of the above proposition verify imposed conditions for MLE consistency of Theorem 5.7 in van der Vaart \cite{deVaart}. Therefore, they can be used to show the consistency of $\widehat{\theta}_0$, see \cite{Frydman2020}.
\begin{theo}\label{theo:maintheo}
Under the regularity condition (\ref{eq:bounded}), $\widehat{\theta}_0\stackrel{\mathbb{P}}{\Longrightarrow} \theta_0$ as $K\rightarrow \infty$. Furthermore,
\begin{eqnarray}\label{eq:maintheo}
\sqrt{K}\big(\widehat{\theta}_0 -\theta_0\big) \sim N\big(0, I_p^{-1}(\theta_0) I_c(\theta_0) I_p^{-1}(\theta_0)\big),
\end{eqnarray}
where $I_c(\theta_0)$ is given in (\ref{eq:expfisher}), whilst $I_p(\theta)$ is the expected Fisher information matrix of $X$, $$I_p(\theta_i,\theta_j)=-\mathbb{E}_{\theta}\Big(\frac{\partial^2}{\partial \theta_i \partial \theta_j} \log f_{\theta}(X)\Big).$$
\end{theo}

\noindent \textbf{Proof:} See Appendix F for details of the derivation. $\blacksquare$

%

\subsection{Asymptotic efficiency of the MLE $\widehat{\theta}_0$}
Asymptotic efficiency of the MLE is shown by comparing the covariance matrix $\Sigma(\theta_0)$ of $\widehat{\theta}_0$ with the asymptotic variance $I_p^{-1}(\theta_0)$, which corresponds to the Cram\'er-Rao lower bound. See p.2162 of Newey and McFadden \cite{Newey} or p.489 of Cram\'er \cite{Cramer}.
\begin{prop} Following \cite{Frydman2020}, the asymptotic variance of the MLE $\widehat{\theta}_0$ is given by
\begin{align*}
I_p^{-1}(\theta_0)=\big[\Sigma(\theta_0) I_c(\theta_0)\big]^{1/2} I_c^{-1}(\theta_0),
\end{align*}
where $\Sigma(\theta_0):=\mathbb{E}_{\theta_0}\Big[\big(\widehat{\theta}_0-\theta_0\big)\big(\widehat{\theta}_0-\theta_0\big)^{\top}\Big]$ is a $(Mp^2\times Mp^2)-$covariance matrix with
\begin{eqnarray}
\mathbb{E}_{\theta_0}\big[\big(\widehat{\phi}_{y,n}^0-\phi_{y,n}^0\big)\big(\widehat{\phi}_{x,m}^0-\phi_{x,m}^0\big)\big]&=&\frac{\phi_{y,n}^0}{\alpha_{y}^0}\delta_x(y)\big(\delta_m(n)-\phi_{x,m}^0\big), \nonumber \\
\mathbb{E}_{\theta_0}\big[\big(\widehat{\phi}_{x^{\prime},n}^0-\phi_{x^{\prime},n}^0\big)\big(\widehat{q}_{xy,m}^0-q_{xy,m}^0\big)\big]&=&0,  \label{eq:CovMat}\\
\mathbb{E}_{\theta_0}\big[\big(\widehat{q}_{x^{\prime}y^{\prime},n}^0-q_{x^{\prime}y^{\prime},n}^0\big)\big(\widehat{q}_{xy,m}^0-q_{xy,m}^0\big)\big]&=&\frac{q_{x^{\prime}y^{\prime},n}^0 \delta_m(n)\delta_x(x^{\prime})\delta_y(y^{\prime})  }{ \alpha_0^{\top}S_n^0 \int_0^T e^{Q_n^0 u} e_x du }. \nonumber
\end{eqnarray}
\end{prop}

\noindent \textbf{Proof:} Following the distribution (\ref{eq:maintheo}) and the result of Theorem 1 of \cite{Frydman2020}, we obtain $$\Sigma(\theta_0)=I_p^{-1}(\theta_0) I_c(\theta_0) I_p^{-1}(\theta_0),$$ from which the claim follows after solving for the matrix $I_p^{-1}(\theta_0)$ in the stated identity. $\blacksquare$

\pagebreak

Comparing the matrices $\Sigma(\theta_0)$ (\ref{eq:CovMat}) and $I_c(\theta_0)$ (\ref{eq:expfisher}), we have $\Sigma(q^0)=I_c^{-1}(q^0)$, i.e., the covariance matrix of $\widehat{q}^0:=(\widehat{q}_{xy,m}^0: (x,y)\in\mathbb{S}, 1\leq m\leq M)$ equals to the inverse of submatrix $I_c(q^0)$. Hence,
\begin{equation}\label{eq:matIp}
\begin{split}
I_p^{-1}(\theta_0)=&\left[
\begin{pmatrix}
\Sigma(\phi^0) & 0 \\ 
0 & \Sigma(q^0)
\end{pmatrix}
\begin{pmatrix}
I_c(\phi^0) & 0 \\ 
0 & I_c(q^0)
\end{pmatrix}
\right]^{1/2}\left(
\begin{array}{cc}
I_c(\phi^0) & 0 \\ 
0 & I_c(q^0)
\end{array}
\right)^{-1}\\
=&\left(
\begin{array}{cc}
\left[\Sigma(\phi^0)I_c(\phi^0)\right]^{1/2} I_c^{-1}(\phi^0) & 0\\
0 & I_c^{-1}(q^0) \\
\end{array}
\right),
\end{split}
\end{equation}
from which $\Sigma(q^0)=I_p^{-1}(q^0)$. Thus, at its convergence, the MLE $\widehat{q}^0$ (\ref{eq:theMLE}) forms an asymptotically efficient estimator of $q^0$. Next, we wanted to investigate the asymptotic efficiency of the MLE $\widehat{\phi}^0$. For simplicity, consider a two-regime switching model with $p$ number of states. Since $\phi_{i,1}^0+\phi_{i,2}^0=1$, the true parameter $\phi^0$ of the regime selection is given by a $(p\times 1)$vector $\phi^0=(\phi_{1,1}^0,\cdots,\phi_{p,1}^0)$. It is straightforward to check following (\ref{eq:CovMat}) and (\ref{eq:expfisher}) that $\Sigma(\phi^0)=\textrm{diag}\Big(\frac{\phi_{1,1}^0}{\alpha_1^0}(1-\phi_{1,1}^0) ,\cdots, \frac{\phi_{p,1}^0}{\alpha_p^0}(1-\phi_{p,1}^0) \Big)$ and  $I_c(\phi^0)=\textrm{diag}\Big(\frac{\alpha_1^0}{\phi_{1,1}^0}(1-\phi_{1,1}^0) ,\cdots, \frac{\alpha_p^0}{\phi_{p,1}^0}(1-\phi_{p,1}^0) \Big)$. Thus, as $0<\alpha_x^0, \phi_{x,m}^0<1$ for all $x\in\mathbb{S}$, $1\leq m\leq M$, the inverse of submatrix $I_p(\phi^0)$ is given following the identity (\ref{eq:matIp}) by
\begin{align*}
I_p^{-1}(\phi^0)=\left[\Sigma(\phi^0)I_c(\phi^0)\right]^{1/2}I_c^{-1}(\phi^0)=\textrm{diag}\Big(\frac{\phi_{1,1}^0}{\alpha_1^0} ,\cdots, \frac{\phi_{p,1}^0}{\alpha_p^0} \Big) \geqq \Sigma(\phi^0),
\end{align*}
which leads to the conclusion that, at the convergence, the covariance matrix $\Sigma(\phi^0)$ of the MLE $\widehat{\phi}^0$ (\ref{eq:theMLE}) is found to be less than the Cram\'er-Rao lower bound $I_p^{-1}(\phi^0)$. Note that $C \geq D$ denotes Loewner ordering of two matrices so that C-D is non-negative definite. For higher number of regime $M\geq 3$, we rely on numerical computation to be discussed in more details in the next section.  

\section{Simulation study}\label{sec:sec6}

This section exemplifies the main results presented in Sections \ref{sec:sec3} - \ref{sec:sec5} through a series of simulation studies. In particular, we are interested in applying the EM algorithm to get maximum likelihood estimates $\widehat{\theta}_0$ of the distribution parameter $\theta_0$ of the process $X$ (\ref{eq:main}) based on Monte Carlo simulation of its sample paths. Algorithm for simulation of the sample paths is presented. Estimated standard error $\widehat{\textrm{Var}}(\widehat{\theta}_0)$ of the MLE $\widehat{\theta}_0$ is obtained using the inverse of observed Fisher information matrix $J_{p}(\widehat{\theta}_0)$ and compared with the actual standard errors $\textrm{Var}(\widehat{\theta}_0)$. Asymptotic properties (consistency, normality, and efficiency) of the MLE $\widehat{\theta}_0$ are verified and are confirmed by the simulation results.   

\subsection{Specification of initial value of distribution parameters $\theta_0$}\label{subsec:initial}

For simulation study, let $\mathbb{S}=\{1,2,3\}$ and $M=3$. The value of initial probabilities $\alpha_x^0$ and $\phi _{x,m}^0$ are presented in Table \ref{table:parvalue}. 
\begin{table}[ht!]
\centering 
\begin{tabular}{ccccc}
\hline\hline
State (x) & $\alpha_x^0$ & $\phi_{x,1}^0$ & $\phi_{x,2}^0$ & $\phi_{x,3}^0$ \\%
[0.5ex] \hline
1 & 1/3  & 0.5 & 0.3  & 0.2\\ 
2 & 1/3  & 0.25 & 0.55 & 0.2\\ 
3 & 1/3 & 0.6  & 0.1 & 0.3\\[1ex] \hline
\end{tabular}
\caption{Parameter values for $\alpha_{x}^0$ and $\phi_{x,m}^0$, $m=1,2,3$. }
\label{table:parvalue}
\end{table}

The intensity matrices $Q_{1}^0$, $Q_{2}^0$, and $Q_3^0$ for the regime membership $X^{(1)}$, $X^{(2)}$, and $X^{(3)}$ are 
\begin{eqnarray*}
Q_{1}^0=\left( 
\begin{array}{ccc}
-2.0 & 1.2 & 0.8 \\ 
0.2 & -0.4 & 0.2 \\ 
1.2 & 1.8 & -3.0%
\end{array}%
\right), \quad Q_{2}^0=\left( 
\begin{array}{ccc}
-3.0 & 2.4 & 0.6 \\ 
0.2 & -0.4 & 0.2 \\ 
0.4 & 1.6 & -2.0%
\end{array}%
\right),
\end{eqnarray*}%
and
\begin{eqnarray*}
Q_{3}^0=\left( 
\begin{array}{ccc}
-4.0 & 1.6 & 2.4 \\ 
0.2 & -0.4 & 0.2 \\ 
3.0 & 2.0 & -5.0%
\end{array}%
\right),
\end{eqnarray*}
respectively. The transition matrices of embedded Markov chains $Z^{(m)},m=1,2,3$ are subsequently
\begin{eqnarray*}
\Pi_{1}^0=[\pi_{xy,1}^0]_{xy}= 
\begin{pmatrix}
0 & 0.6 & 0.4 \\ 
0.5 & 0 & 0.5 \\ 
0.4 & 0.6 & 0%
\end{pmatrix}, \quad \Pi_{2}^0=[\pi_{xy,2}^0]_{xy}= 
\begin{pmatrix}
0 & 0.8 & 0.2 \\ 
0.5 & 0 & 0.5 \\ 
0.2 & 0.8 & 0%
\end{pmatrix},
\end{eqnarray*}%
and
\begin{eqnarray*}
\Pi_{3}^0=[\pi_{xy,3}^0]_{xy}= 
\begin{pmatrix}
0 & 0.4 & 0.6 \\ 
0.5 & 0 & 0.5 \\ 
0.6 & 0.4 & 0%
\end{pmatrix}.
\end{eqnarray*}
We see following the matrices $Q_m^0$ and $\Pi_m^0$, $m=1,2,3$, that each regime $X^{(m)}$ has different expected state occupation time and the probability of making a jump from one state to another, except for the transition from state 2. In the latter case, it is difficult to identify which underlying process that drives dynamics of $X$ (\ref{eq:main}) when it moves from state 2. This adds complexity in the estimation. However, due to consistency of the MLE (Theorem \ref{theo:maintheo}), the EM algorithm provides accurate estimation for the transition rate $(q_{21,m},q_{23,m})$ as we can see from Tables \ref{table:biasmse} and \ref{table:eststdev}. 

\subsection{Algorithm for the simulation of sample paths}\label{sec:algol}
Sample paths $\{X_n,T_n: T_n\leq T\}$ of $X$ (\ref{eq:main}) on $[0,T]$, with $T<\infty$, are generated based on iterations (\ref{eq:Xn})-(\ref{eq:Phin}). The algorithm is given below.

\medskip

\begin{enumerate}

\item[(i)][\textbf{Initial time}] Let $\mathbb{S}=\{1,2,\cdots,p\}$ and $t_0=0$. Set $T_0=t_0$ and specify the number of regime $M\geq 1$, and initial parameter value $\theta_0=(\phi_{x,m}^0, q_{xy,m}^0: (x,y)\in\mathbb{S}, 1\leq m\leq M)$. Calculate the transition probability $\pi_{xy,m}^0:=q_{xy,m}^0/q_{x,m}^0$ with $q_{x,m}^0=\sum_{y\neq x, y\in \mathbb{S}} q_{xy,m}^0$.

\medskip

\begin{enumerate}

\item Draw independently uniform random variates $U_0,V_0,W_0\sim U[0,1]$.

\medskip

\item Select an initial state $X_0=x_0\in\mathbb{S}$ using $\boldsymbol{\alpha}_0$ as such that $\sum_{i=1}^{x_0-1} \alpha_i^0 < U_0 \leq \sum_{i=1}^{x_0} \alpha_i^0$. 

\medskip

\item Use $\phi_{x_0,\bullet}^0$ to choose a regime $\Phi_0=m$ as such that $\sum_{\ell=1}^{m-1} \phi_{x_0,\ell}^0< W_0 \leq \sum_{\ell=1}^{m} \phi_{x_0,\ell}^0$.

\end{enumerate}

\medskip

\item[(ii)][\textbf{Step-1}] Draw independently uniform random variates $U_1,V_1,W_1\sim U[0,1]$. 

\medskip

\begin{enumerate}
\item Determine the epoch time $T_1=t_1$ with $t_1=t_0-\log V_0/q_{x_0,\Phi_0}^0$. 

\medskip

\item Find $X_1=x_1$ as such that $\sum_{w=1}^{x_1-1}\pi_{x_0w,\Phi_0}^0 < U_1\leq \sum_{w=1}^{x_1}\pi_{x_0w,\Phi_0}^0$. 

\medskip

\item Based on the observation $\mathcal{H}_{t_1,x_1}=(x_0,t_0,x_1,t_1)$, calculate the statistics (\ref{eq:stats}): $B_{x}$, $N_{xy}$, and $T_x$ for all $(x,y)\in \mathbb{S}$. 

\medskip

\item Use these statistics to find using the formula (\ref{eq:phixt}) the conditional switching probability $\phi_{x_1,m}(t_1)=\mathbb{P}\{\Phi=m\vert \mathcal{H}_{t_1,x_1}\}$ for $1\leq m\leq M$.

\medskip

\item Then select the regime membership $\Phi_1=m$ as such that $\sum_{\ell=1}^{m-1} \phi_{x_1,\ell}(t_1) < W_1 \leq \sum_{\ell=1}^{m} \phi_{x_1,\ell}(t_1)$. The selected regime $\Phi_1$ is used to determine $T_2$ and $X_2$. 

\end{enumerate}

\medskip

\item[(iii)][\textbf{Step-(n+1)}] Suppose that iterations (\ref{eq:Xn})-(\ref{eq:Phin}) have been repeated $n$ times so that we have generated an observation $\mathcal{H}_{t_n,x_n}=\{x_0,t_0,x_1,t_1,\cdots,x_n,t_n\}$ of the sample paths and a sequence of independent uniform random variates $(U_k,V_k,W_k:k=1,\cdots,n)$. 

\medskip

\begin{enumerate}

\item Based on the observation $\mathcal{H}_{t_n,x_n}$, calculate the statistics $B_{x}$, $N_{xy}$, and $T_x$ $\forall (x,y)\in \mathbb{S}$. 

\medskip

\item Use these statistics to find using the formula (\ref{eq:phixt}) the conditional switching probability $\phi_{x_n,m}(t_n)=\mathbb{P}\{\Phi=m\vert \mathcal{H}_{t_n,x_n}\}$ for $1\leq m\leq M$. 

\medskip

\item Then select the regime membership $\Phi_n=m$ as such that $\sum_{\ell=1}^{m-1} \phi_{x_n,\ell}(t_n) < W_n \leq \sum_{\ell=1}^{m} \phi_{x_n,\ell}(t_n)$. 

\medskip

\item Set $T_{n+1}=t_{n+1}$ with $t_{n+1}=t_n-\log V_n/q_{x_n,\Phi_n}^0$. 

\medskip

\item Draw an independent uniform random variate $U_{n+1}\sim U[0,1]$ and select $X_{n+1}=x_{n+1}\in\mathbb{S}$ as such that $\sum_{w=1}^{x_{n+1}-1}\pi_{x_n w,\Phi_n}^0 < U_{n+1}\leq \sum_{w=1}^{x_{n+1}}\pi_{x_n w,\Phi_n}^0$.

\end{enumerate}

\medskip

\item[(iv)][\textbf{Stopping criterion}] Stop if $t_{n+1}>T$. Otherwise, increase $n$ and go back to (iii).

\end{enumerate}

\subsection{Simulation and estimation results}

Based on the above parameters, a specified $K$ independent sample paths of $X$ (\ref{eq:main}) are generated using the algorithm. Figure \ref{paths} exhibits two generated sample paths of $X$ along with the plots of their regime membership. We observe from the sample paths (A) and (C) that $X$ may change the regime more frequently compared to paths (B) and (D). 

Using these $K$ independent observations, the EM algorithm is applied to obtain maximum likelihood estimate $\widehat{\theta}_0$ of the initial value $\theta_0$. To verify the consistency of the MLE $\widehat{\theta}_0$, a set of $N=200$ independent sample paths of size $K=750, 1500,$ and $K=3000$ are generated respectively. To each set $N=200$ observations of a specified sample size $K$, the EM algorithm is applied to get $N$ independent sets of MLE $\{\widehat{\theta}_{n,K}:n=1,\cdots,N\}$. This process is repeated for each $K$. The biases $N^{-1}\sum_{n=1}^N (\widehat{\theta}_{n,K}-\theta_0)$ and estimated standard errors (SE) of $\widehat{\theta}_0$ are calculated using the inverse of information matrix $J_p(\widehat{\theta}_0)$, compared with the root of mean squared error (RMSE). Furthermore, the p-value of the Kolmogorov-Smirnov statistic (KS) is computed for the standardized biases. The KS statistic provides a goodness-of-fit measure between the empirical distribution of the biases and the $N(0,1)$ cumulative distribution function. The results are presented in Table \ref{table:biasmse} and Table \ref{table:eststdev}. 

Table \ref{table:biasmse} shows that both Bias($\widehat{\theta}_0$) and RMSE($\widehat{\theta}_0$) decrease by the sample size $K$, which in turn verifies the consistency of the MLE $\widehat{\theta}_0$. Table \ref{table:eststdev} presents the parameter estimate $\widehat{\theta}_0=\frac{1}{N}\sum_{n=1}^N \widehat{\theta}_{n,K}$, for $K=3000$ and $N=200$, along with its estimated SE $\sqrt{J_p^{-1}(\widehat{\theta}_0)}$ and the actual SE $\sqrt{\textrm{Var}(\widehat{\theta}_0)}$. The table shows the accuracy of the MLE $\widehat{\theta}_0$ as it is very close to the initial value $\theta_0$ with estimated SE also very close to the actual one. The p-value of KS statistic is larger than $5\%$. Hence, the asymptotic normality of the biases is statistically significant at the acceptance level $\alpha=5\%$.

To verify the asymptotic efficiency of the MLE $\widehat{\phi}^0$, the covariance matrix $\Sigma(\phi^0)$ of $\sqrt{K}\big(\widehat{\phi}^0-\phi^0\big)$ is compared to the Cram\'er-Rao lower bound $I_p^{-1}(\phi^0)=\left[\Sigma(\phi^0)I_c(\phi^0)\right]^{1/2}I_c^{-1}(\phi^0)$. Using the result of Theorem 1 in \cite{VanLoan}, the two matrices are given below. The result clearly shows that $\Sigma(\phi^0)\leq I_p^{-1}(\phi^0)$ which confirms that the asymptotic variance of $\widehat{\phi}^0$ is less than its Cram\'er-Rao lower bound.

\begin{align*}
\Sigma(\phi^0)=
\begin{pmatrix}
0.75 & -0.45 & 0 & 0 & 0 & 0\\
-0.45 & 0.63 & 0 & 0 & 0 & 0\\
0 & 0 & 0.5625 & -0.4125 & 0 & 0\\
0 & 0 & -0.4125 & 0.7425 & 0 & 0\\
0 & 0 & 0 & 0 & 0.72 & -0.18\\
0 & 0 & 0 &0 & -0.18 & 0.27
\end{pmatrix},
\end{align*}
and
\begin{align*}
I_p^{-1}(\phi^0)=
\begin{pmatrix}
5.25 & 2.25 & 0 & 0 & 0 & 0\\
2.25 & 2.25 & 0 & 0 & 0 & 0\\
0 & 0 & 1.6875 & 2.0625 & 0 & 0\\
0 & 0 & 2.0625 & 6.1875 & 0 & 0\\
0 & 0 & 0 & 0 & 5.4 & 0.6\\
0 & 0 & 0 &0 & 0.6 & 0.4
\end{pmatrix}.
\end{align*}

\begin{table}[t!]
\centering
\begin{tabular}{|l|l|l|l|l|l|l|l|l|l|}
\hline
\multirow{2}{*}{$\;\;\theta_0$} & \multicolumn{1}{c|}{True} & \multicolumn{3}{c|}{Bias ($10^{-2}$)} & \multicolumn{3}{c|}{RMSE ($10^{-2} $)}   \\ 
              & \,\textrm{Value} & $K=750$ & $K=1500$ & $K=3000$ & $K=750$ & $K=1500$ & $K=3000$  \\ \hline
$\phi_{1,1}$ &  0.5 & -0.58256 & -0.35108 & -0.06198 & 4.92557 & 3.18217 & 2.31614 \\ 
$\phi_{1,2}$ &   0.3 & 0.69808 & 0.43228 & 0.00009 & 4.33558 & 2.95272 & 2.13886 \\ 
 $\phi_{2,1}$ &  0.25 & 0.32694 & -0.23725 & 0.03407 & 5.33382 & 2.98005 & 2.24124 \\ 
  $\phi_{2,2}$ & 0.55 & -2.29103 & 0.29762 & 0.07108 & 9.63232 & 4.67092 & 2.24844 \\ 
$\phi_{3,1}$ &  0.6 & -1.19392 & 0.24058 & -0.09421 & 6.74317 & 2.84524 & 1.98517 \\ 
 $\phi_{3,2}$ & 0.1 & 1.74463 & 0.26530 &  -0.00244 & 7.87991 & 2.64431 & 1.46116 \\ 
 $q_{12,1}$ & 1.2 & 0.87180 & -0.34526 & -0.04485 & 11.55583 & 2.76056 & 1.83873 \\ 
 $q_{13,1}$ & 0.8 & -0.49850 & 0.24439 & -0.00149 & 3.10226 & 2.09725 & 1.44143 \\ 
 $q_{21,1}$ & 0.2 & -0.02958 & 0.02796 & -0.00955 & 0.58126 & 0.42424 & 0.27316 \\ 
$q_{23,1}$ & 0.2 & -0.03884 & 0.01506 & -0.00076 & 0.55036 & 0.42062 & 0.28809 \\ 
  $q_{31,1}$ & 1.2 & -0.39414 & 0.38352 & -0.10264 & 9.11007 & 3.20518 & 2.34497 \\ 
  $q_{32,1}$ & 1.8 & 0.28056 & 0.10690 & 0.05345 & 4.88136 & 3.55120 & 2.45693 \\ 
  $q_{12,2}$ & 2.4 & -5.99395 & -0.78335 & 0.33378 & 22.92851 & 10.55531 & 4.63466 \\ 
 $q_{13,2}$ & 0.6  & 7.99767 & 1.70836 & 0.01445 & 37.68517 & 17.94271 & 2.10879 \\ 
  $q_{21,2}$ & 0.2& 0.05800 & -0.03413 & 0.00371 & 0.61519 & 0.44973 & 0.29654 \\ 
  $q_{23,2}$ & 0.2 & -0.12517 & -0.05206 & -0.00214 & 0.66612 & 0.47188 & 0.30304 \\ 
   $q_{31,2}$ & 0.4 & 12.13783 & 2.58894 & -0.06130 & 54.60797 & 25.97195 & 1.54233 \\ 
 $q_{32,2}$ & 1.6 & 1.41371 & 0.93268 & -0.06946 & 10.49528 & 5.73001 & 2.44677 \\ 
   $q_{12,3}$ & 1.6 & 3.39961 & 1.46223 & -0.29979 & 18.88245 & 8.92913 & 2.74420 \\ 
  $q_{13,3}$ & 2.4 & -7.09454 & -1.29760 & 0.24680 & 38.85252 & 18.68112 & 3.70443 \\ 
 $q_{21,3}$ & 0.2 & 0.11092 & 0.05177 & 0.04362 & 0.70087 & 0.50511 & 0.38840 \\ 
  $q_{23,3}$ & 0.2 & 0.08326 & -0.07597 & 0.04897 & 0.73597 & 0.46801 & 0.34695 \\ 
   $q_{31,3}$ & 3.0 & -11.60731 & -2.87170 & -0.23894 & 56.24500 & 26.59106 & 4.64256 \\ 
  $q_{32,3}$ & 2.0 & -1.01723 & -0.76809 & -0.136821 & 11.16507 & 6.46066 & 3.35442 \\ 
   \hline
\end{tabular}
\caption{Parameter $\theta_0$, its true value, Bias($\widehat{\theta}_0$)=$\frac{1}{N}\sum\limits_{n=1}^N \big(\widehat{\theta}_{n,K}-\theta_0\big)$ and RMSE($\widehat{\theta}_0$)
=$\Big[\frac{1}{N}\sum\limits_{n=1}^N \big(\widehat{\theta}_{n,K} -\theta_0\big)^2\Big]^{1/2}$, with fixed $N=200$, for different size $K$ of the sample paths $\{X^k:k=1,\cdots,K\}$. The EM iteration stops when $\Vert \widehat{\theta}_0^{\ell+1} - \widehat{\theta}_0^{\ell}\Vert < 10^{-5}$. }
\label{table:biasmse} 
\end{table}

\begin{table}[t!]
{\centering 
\par
\begin{tabular}{|l|l|l|l|l|l|l|l|l|l|l|}
\hline
\multirow{2}{*}{$\;\;\theta_0$} & \multicolumn{1}{c|}{True} & \multicolumn{1}{c|}{Estimate} & \multicolumn{2}{c|}{Standard Error (\%)}  & \multicolumn{1}{c|}{KS} \\ 
& \,\,\textrm{Value} & $\;\;\;\;\;\widehat{\theta}_0$ & $\sqrt{\text{Var}(%
\widehat{\theta}_0)}$ & $\sqrt{J^{-1}(\widehat{\theta}_0)}$ &  \,\,\textrm{Test} \\ \hline
$\phi_{1,1}$ &  0.5  & 0.49938 & 2.31531  & 2.71894  & 0.9995\\ [1pt]
$\phi_{1,2}$ &   0.3 & 0.30000 & 2.13886 & 2.76167  & 0.9939 \\ [1pt]
$\phi_{2,1}$ &  0.25 & 0.25034 & 2.24098  & 2.24678   & 0.9025 \\  [1pt]
$\phi_{2,2}$ & 0.55 & 0.55071 &2.24732  & 2.76028  &  0.8244\\  [1pt]
$\phi_{3,1}$ &  0.6 & 0.59906 & 1.98293 & 2.22130    & 0.9386\\  [1pt]
$\phi_{3,2}$ & 0.1 & 0.09998 &1.46116  & 1.45802   &  0.9294 \\  [1pt]
$q_{12,1}$ & 1.2  & 1.19955 &1.83818  & 1.84434   & 0.6437 \\ [1pt]
$q_{13,1}$ & 0.8 & 0.79999 &1.44143  &  1.39853    & 0.9015 \\  [1pt]
 $q_{21,1}$ & 0.2 & 0.19999 & 0.27299 & 0.28991   & 0.8788 \\  [1pt]
$q_{23,1}$ & 0.2 & 0.19999 & 0.28809    & 0.28966  & 0.9502\\  [1pt]
$q_{31,1}$ & 1.2 & 1.19897 & 2.34272  & 2.34950   & 0.2966\\  [1pt]
$q_{32,1}$ & 1.8 & 1.80053 & 2.45635   & 2.47790  & 0.6235 \\  [1pt]
$q_{12,2}$ & 2.4 & 2.40334 &4.62263   & 4.69439  & 0.951\\  [1pt]
$q_{13,2}$ & 0.6 & 0.60014 &2.10874  &  2.06770   & 0.9022\\  [1pt]
$q_{21,2}$ & 0.2  & 0.20004 &0.29652  &  0.33264  &0.6413 \\  [1pt]
 $q_{23,2}$ & 0.2 & 0.19998 &0.30303 & 0.33279  & 0.8133\\  [1pt]
 $q_{31,2}$ & 0.4 & 0.39939  &1.54111  & 1.55721  & 0.3478\\ [1pt]
$q_{32,2}$ & 1.6  & 1.59931 & 2.44578   & 2.63274   & 0.8322  \\ [1pt]
$q_{12,3}$ & 1.6  & 1.59700 & 2.72778    & 2.84104   &0.7401   \\  [1pt]
$q_{13,3}$ & 2.4 & 2.40247 & 3.69620    & 3.72896   & 0.9839   \\  [1pt]
$q_{21,3}$ & 0.2 & 0.20044 & 0.38594    & 0.36897     &0.8922  \\ [1pt]
$q_{23,3}$ & 0.2 & 0.20049 &0.34348    & 0.36875    & 0.9728  \\ [1pt]
$q_{31,3}$ & 3.0 & 2.99761 &4.63641   & 4.56372    &0.9985   \\  [1pt]
$q_{32,3}$ & 2.0 & 1.99632 &3.33415   & 3.52475   & 0.7301    \\  \hline
\end{tabular}
\caption{MLE $\widehat{\theta}_0=\frac{1}{N}\sum_{n=1}^N \widehat{\theta}_{n,K}$, estimated standard errors of $\widehat{\theta}_0$ using $\textrm{Var}(\widehat{\theta}_0)=\textrm{RMSE}(\widehat{\theta}_0)^2 -\textrm{Bias}(\widehat{\theta}_0)^2$ and the inverse of $J(\widehat{\theta}_K)=\frac{1}{N}\sum_{n=1}^N I(\widehat{\theta}_{n,K})$, for $K=3000$ and $N=200$. The last column lists the p-value of Kolmogorov-Smirnov
statistic for goodness-of-fit between empirical CDF of standardized biases
and N(0,1) CDF.}
\label{table:eststdev}
}
\end{table}

\subsection{Model selection}
To further verify the accuracy of the EM algorithm in terms of the number of regime membership $M\geq 1$, the following simulation study was performed. First, $K=10000$ independent sample paths of $X$ (\ref{eq:main}) with initial $M=2$ regime memberships were generated based on the same initial distribution $\alpha_0$ and the intensity matrices $Q_1^0$ and $Q_2^0$. However, the regime membership probability $\phi^0=(\phi_{1,1}^0,\phi_{2,1}^0,\phi_{3,1}^0)$ is set to be $(0.5,0.25,0.6)$. Assuming the initial value of $M$ is unknown, the EM algorithm is repeatedly applied to these realized sample paths for different values of $M$ ranging from $M=1$ (Markov model) and $M=6$. To strike the balance between the model's fit and its complexity in terms of the number of model's parameters, Akaike information criterion \cite{Akaike} is used. In our notation, $\textrm{AIC}_M=2\vert \theta_0^M\vert -2\log\mathcal{L}(\widehat{\theta}_0^M)$, with $\mathcal{L}(\theta_0^M)= f_{\theta_0^M}(\mathcal{X})$ and $\vert \theta_0^M\vert$ representing the dimension of parameter $\theta_0^M$ for $M$-regime. The best fitted model is the one with the smallest $\textrm{AIC}_M$. Secondly, repeat the simulation study for initial $M=3$ regime membership with the same initial distributions $\alpha_0$, $\phi_0$, and the same intensity matrices $Q_1^0$, $Q_2^0$, and $Q_3^0$ as specified in Section 6.1. Compute $\textrm{AIC}_M$ for each fitted model. Table \ref{table:AICBIC2} summarizes $\textrm{AIC}_M$ and the log-likelihood $\log \mathcal{L}(\widehat{\theta}_0^M)$ for each simulation study. The table shows that $\log \mathcal{L}(\widehat{\theta}_0^M)$ increases by the number of regime $M$. However, the $\textrm{AIC}_M$ has its smallest value \textbf{729952} for 2-regime under the null hypothesis $H_0: M=2$, and \textbf{717864.5} for 3-regime under $H_0: M=3$. The results show that the EM algorithm specifies accurately the number of regime in the generated sample paths.

\begin{table}[h!]
{\centering 
\begin{tabular}{|l|l|l|l|l|l|l|l|l|l||l|}
\hline
 \multicolumn{1}{|c|}{Fitted} & \multicolumn{2}{c|}{$H_0$: M=2} & \multicolumn{2}{c|}{$H_0$: M=3}   \\ 
 \;\;\;\textrm{Model}                & \;\; $\textrm{AIC}_M$ & \;$\log \mathcal{L}(\widehat{\theta}_0^M)$ & \;\;\; $\textrm{AIC}_M$ & \; $\log \mathcal{L}(\widehat{\theta}_0^M)$  \\[0.5ex] \hline
Markov & 735568.1 & -367776 &   739572.2 & -369778.1    \\[1pt] 
2 Regime & \textbf{729952} & -364959   &   721979.3 & -360972.6   \\[1pt] 
3 Regime & 729956.8 & -364952.4 &   \textbf{717864.5} & -358906.2 \\[1pt] 
4 Regime & 729960.7 & -364945.3  &  717875.3 & -358902.6 \\[1pt] 
5 Regime & 729968 & -364940  &  717883.8 & -358897.9 \\[1pt] 
6 Regime & 729973.8 & -364933.9  &  717888.4 & -358891.2 \\[1ex] \hline
\end{tabular}
\caption{Summary of model statistics $\textrm{AIC}_M$ and $\log \mathcal{L}(\widehat{\theta}_0^M)$.}
\label{table:AICBIC2}
}
\end{table}

\section{Concluding Remarks}\label{sec:sec7}

This paper proposed and estimated a new class of conditional Markov jump processes with regime switching and path dependence. It is a non-trivial generalization of the Markov process which allows the process to adjust the transition rate as it moves from one state to another based on its current state and time as well as its past trajectories through their likelihood function. The transition from current state to another state depends on the amount of time the process stays in the state. It was shown that the process has a distributional equivalent stochastic representation with a general mixture of Markov jump processes discussed in \cite{Frydman2020} and \cite{Surya2020}. Maximum likelihood estimates (MLE) of the distribution parameters of the process are given in closed form which makes the EM computation for the estimation fast and stable. Asymptotic properties (consistency, asymptotic normality, and efficiency) of the MLE were derived. In particular, the covariance matrix of the MLE for transition rates coincides with the Cram\'er-Rao lower bound (the inverse of expected Fisher information matrix), and is less for the covariance matrix of the MLE for regime membership. These findings show that the MLE of the transition rates is \textit{asymptotically efficient}, and is \textit{asymptotically superefficient} for  the MLE of the regime membership. See van der Vaart \cite{deVaart97} for details on supperefficient estimator. Estimated standard errors of the MLE were presented explicitly in a simplified form of Louis' general matrix formula \cite{Louis}. The results complement the recent work on maximum likelihood estimation for the general mixture of Markov jump processes \cite{Frydman2020}.  A series of simulation study show that the estimation was accurate and confirmed the asymptotic properties of the MLE. For future works, we intend to estimate the (joint) probability distribution of exit times to absorbing states of the proposed conditional Markov jump process using real dataset and covariates under periodic observation of the sample paths. Its non-stationary property, ability to explicitly include past information in the distribution, and available explicit maximum likelihood estimates should offer potential for variety of applications of the model.  

\section*{Acknowledgements}
This work was motivated by the author's recent joint work with Professor Halina Frydman of New York University Stern School of Business whom he visited in September 2019. He thanks Professor Frydman for the invitation, support, valuable discussions, and hospitality provided during his stay at the NYU Stern. 
The author also acknowledges financial support from the School of Mathematics and Statistics of Victoria University of Wellington for the research grant \# 218772.

\appendix

\section{Proof of Proposition \ref{prop:prop1}}
On recalling from (\ref{eq:Tn}), the sequence of epoch times $\{T_n\}_{n\geq 0}$ of $X$ can be singled out as 
\begin{align}\label{eq:recTn}
T_n=T_0-\sum_{k=0}^{n-1}\sum_{m=1}^M \frac{\log V_k}{q_{X_k,m}}\delta_{m}\big(\Phi(X_k,T_x,W_k)\big), \quad \textrm{for \; $n\geq 1$},
\end{align}
where $X_k=X(T_k)$ is the state reached at epoch time $T_k$.
Given past observation $\mathcal{H}_{t-}$ of the sample paths of $X$, and that $X$ starts in state $x\in\mathbb{S}$ at time $t\geq 0$, let $T_0=t_0=t$ and $X_0=X(T_0)=x$. Thus, $\tau_t \stackrel{d}{=} T_1$ under $\mathbb{P}\{\bullet\vert \mathcal{H}_{t,x}\}$. Therefore, by the law of total probability and the Bayes formula, 
\begin{align*}
\mathbb{P}\big\{\tau_t>r \big\vert X(t)=x,\mathcal{H}_{t-}\big\}=&\mathbb{P}\big\{T_1>r \big\vert X(t)=x, \mathcal{H}_{t-}\big\}\\=&\mathbb{P}\Big\{T_0-\sum_{\ell=1}^M \frac{\log V_0}{q_{X_0,\ell}}\delta_{\ell}\big(\Phi(X_0,T_0,W_0)\big) > r\Big\vert X_0=x, T_0=t_0, \mathcal{H}_{t_0-}\Big\} \\
&\hspace{-4cm}=\sum_{m=1}^M \mathbb{P}\Big\{t-\sum_{\ell=1}^M \frac{\log V_0}{q_{x,\ell}}\delta_{\ell}\big(\Phi(x,t,W_0)\big) > r, \Phi(x,t,W_0)=m\Big\vert X_0=x, T_0=t_0, \mathcal{H}_{t_0-}\Big\} \\
&\hspace{-4cm}=\sum_{m=1}^M \mathbb{P}\Big\{\Phi(x,t,W_0)=m\Big\vert X_0=x,T_0=t_0,\mathcal{H}_{t_0-} \Big\} \\
&\hspace{-2.5cm}\times \mathbb{P}\Big\{t-\sum_{\ell=1}^M \frac{\log V_0}{q_{x,\ell}}\delta_{\ell}\big(\Phi(x,t,W_0)\big) > r\Big\vert \Phi(x,t,W_0)=m, X_0=x, T_0=t_0, \mathcal{H}_{t_0-}\Big\} \\
&\hspace{-4cm}=\sum_{m=1}^M \mathbb{P}\Big\{\Phi(x,t,W_0)=m\Big\vert X_0=x,T_0=t_0,\mathcal{H}_{t_0-} \Big\} \\
&\hspace{-2.5cm}\times \mathbb{P}\Big\{-\frac{\log V_0}{q_{x,m}} > r-t\Big\vert \Phi(x,t,W_0)=m, X_0=x, T_0=t_0, \mathcal{H}_{t_0-}\Big\}, 
\end{align*}
from which the identity (\ref{eq:id1}) follows on account that $W_0$ and $V_0$ are independent uniform random variables. Similarly, as the event $\{\tau_t\leq u, X(\tau_t)=y\}\stackrel{d}{=}\{T_1 \leq u, X_1=y\}$ under $\mathbb{P}\{\bullet\vert \mathcal{H}_{t,x}\}$,
\begin{align*}
&\mathbb{P}\big\{\tau_t\leq u, X(\tau_t)=y\big\vert X(t)=x, \mathcal{H}_{t-}\big\}=\mathbb{P}\big\{T_1 \leq u, X_1=y\big\vert X_0=x, T_0=t_0, \mathcal{H}_{t_0-}\big\}\\
&\hspace{0cm}= \mathbb{P}\Big\{ t-\sum_{\ell=1}^M \frac{\log V_0}{q_{x,\ell}}\delta_{\ell}\big(\Phi(x,t,W_0)\big)\leq u, F(x,\Phi(x,t,W_0),U_1)=y\Big\vert X_0=x, T_0=t_0, \mathcal{H}_{t_0-}\Big\}\\
&\hspace{0cm}= \sum_{m=1}^M \mathbb{P}\Big\{ t-\sum_{\ell=1}^M \frac{\log V_0}{q_{x,\ell}}\delta_{\ell}\big(\Phi(x,t,W_0)\big)\leq u, F(x,\Phi(x,t,W_0),U_1)=y, \\
&\hspace{6cm} \Phi(x,t,W_0)=m\Big\vert X_0=x, T_0=t_0, \mathcal{H}_{t_0-}\Big\}\\
&\hspace{0cm}= \sum_{m=1}^M \mathbb{P}\Big\{\Phi(x,t,W_0)=m\Big\vert X_0=x,T_0=t_0, \mathcal{H}_{t_0-}\Big\}\\
&\hspace{2cm}\times \mathbb{P}\Big\{-\frac{\log V_0}{q_{x,m}}\leq u -t, F(x,m,U_1)=y \Big\vert \Phi(x,t,W_0)=m, X_0=x,T_0=t_0, \mathcal{H}_{t_0-}\Big\}\\
&\hspace{0cm}= \sum_{m=1}^M \mathbb{P}\Big\{\Phi(x,t,W_0)=m\Big\vert X_0=x,T_0=t_0, \mathcal{H}_{t_0-}\Big\} \mathbb{P}\Big\{-\frac{\log V_0}{q_{x,m}}\leq u -t\Big\}
\mathbb{P}\big\{F(x,m,U_1)=y \big\},
\end{align*}
leading to the identity (\ref{eq:id2}) on account of $W_0$, $V_0$, and $U_1$ being independent of $X$. $\blacksquare$

\section{Proof of Theorem \ref{theo:transmix}}

Let us first derive an explicit form for the transition probability $P_{xy}^{(m)}(t)$  for $(x,y)\in\mathbb{S}$, with $y\neq x$, $1\leq m\leq M$, and $t\geq 0$. Since $\{U_n\}_{n\geq 0}$ is sequence of independent uniform random variables,
\begin{align}\label{eq:indep}
\mathbb{P}\big\{X_{n+1}=y\big\vert \Phi_n=m,X_n=x, \mathcal{H}_{T_n-}\big\}=&\mathbb{P}\big\{F(X_n,\Phi_n,U_{n+1})=y\big\vert \Phi_n=m,X_n=x, \mathcal{H}_{T_n-}\big\} \nonumber \\
=&\mathbb{P}\big\{F(x,m,U_{n+1})=y\big\}\\
=&\pi_{xy,m}. \nonumber
\end{align}
This is to say that conditional on knowing the current state $X_n$, and all the succession of states $\mathcal{H}_{T_n-}$, the process $X$ moves under $X^{(m)}$ from states $x$ to $y$ independently of the observation $\mathcal{H}_{T_n-}$.

It follows from (\ref{eq:main})-(\ref{eq:Phin}) and applying the Bayes' formula and the law of total probability,
\begin{align}
P_{xy}^{(m)}(r):=&\mathbb{P}\Big\{X(r)=y \Big\vert \Phi_0=m, X_0=x,T_0=0\Big\} \nonumber\\
&\hspace{-2.5cm}=\sum_{n=0}^{\infty} \mathbb{P}\Big\{X_{n}=y,T_{n}\leq r < T_{n+1}\Big\vert \Phi_0=m, X_0=x, T_0=0\Big\} \nonumber\\
&\hspace{-2.5cm}=\sum_{n=0} \sum_{j_1,\cdots,j_{n-1}}\;\; \idotsint\limits_{\{t_1<\cdots< t_{n-1}\}}\; \int_{t_n=t_0}^r \mathbb{P}\Big\{ - \sum_{\ell=1}^M \frac{\log V_{n}}{q_{y,\ell}}\delta_{\ell}\big(\Phi(y,t_n,W_{n})\big) >r-t_n \nonumber \\
&\hspace{4cm}\Big\vert \Phi_0=m, X_n=y, T_n=t_n,\cdots, X_0=x,T_0=t_0\Big\} \nonumber \\
&\hspace{-0.5cm}\times \mathbb{P}\Big\{X_{n}=y, t_0 - \sum_{k=0}^{n-1}\sum_{\ell=1}^M \frac{\log V_{k}}{q_{j_{k},\ell}}\delta_{\ell}\big(\Phi(j_{k},t_{k},W_{k})\big) \in d t_n  \nonumber\\
&\hspace{2.5cm}\Big\vert  \Phi_0=m, X_{n-1}=j_{n-1},T_{n-1}=t_{n-1},\cdots,X_0=x,T_0=t_0 \Big\}\nonumber\\ 
&\hspace{-0.5cm}\times\mathbb{P}\Big\{X_k=j_k, T_k \in dt_k, k=1,\cdots,n-1\Big\vert \Phi_0=m, X_0=x,T_0=t_0\Big\}, \label{eq:trans1}
\end{align}
with $j_0=x, t_0=0$. Since $X(t)=x$, let $T_0=t_0=t\geq 0$ and, hence, $X_0=X(T_0)=x$. Thus, 
\begin{align}
&\mathbb{P}\{X(r)=y, t\leq T_1\leq r \big\vert X(t)=x,\mathcal{H}_{t-}\} \quad \textrm{for $r\geq t$, $(x,y)\in\mathbb{S}$}  \nonumber \\
&\hspace{0cm}=\sum_{w\neq x} \mathbb{P}\big\{X(r)=y, t\leq T_1\leq r, X_1=w \big\vert X(t)=x,\mathcal{H}_{t-}\big\}  \nonumber \\
&\hspace{0cm}=\sum_{w\neq x} \mathbb{P}\Big\{X_r=y, t\leq t_0 -\sum_{\ell=1}^M \frac{\log V_0}{q_{x,\ell}}\delta_{\ell}(\Phi(x,t_0,W_0))\leq r,  \nonumber \\
&\hspace{3cm} F(x,\Phi(x,t_0,W_0),U_1)=w \Big\vert X_0=x,T_0=t_0, \mathcal{H}_{t_0-}\Big\} \nonumber \\
&\hspace{0cm}=\sum_{w\neq x} \sum_{m=1}^M\mathbb{P}\Big\{X(r)=y, t\leq t_0 -\sum_{\ell=1}^M \frac{\log V_0}{q_{x,\ell}}\delta_{\ell}(\Phi(x,t_0,W_0))\leq r,  \nonumber \\
&\hspace{3cm} F(x,\Phi(x,t_0,W_0),U_1)=w,  \Phi(x,t_0,W_0)=m\Big\vert X_0=x,T_0=t_0, \mathcal{H}_{t_0-}\Big\}  \nonumber \\
&\hspace{0cm}=\sum_{w\neq x} \sum_{m=1}^M \int_0^{r-t}\mathbb{P}\Big\{X(r)=y\Big\vert \Phi(x,t,W_0)=m, -\frac{\log V_0}{q_{x,m}}=\nu, X_1=w, X_0=x, T_0=t_0,\mathcal{H}_{t_0-}\Big\}\nonumber \\
&\hspace{2cm}\times\mathbb{P}\Big\{ -\frac{\log V_0}{q_{x,m}}\in d\nu, F(x,m,U_1)=w\Big\vert \Phi(x,t,W_0)=m,X_0=x,T_0=t_0,\mathcal{H}_{t_0-}\Big\} \nonumber\\
&\hspace{2cm}\times \mathbb{P}\Big\{\Phi(x,t,W_0)=m\Big\vert X_0=x,T_0=t_0,\mathcal{H}_{t_0-}\Big\}  \nonumber \\
&=\sum_{m=1}^M \sum_{w\neq x}\int_0^{r-t} \mathbb{P}\Big\{X(r)=y\Big\vert \Phi(x,t,W_0)=m, -\frac{\log V_0}{q_{x,m}}=\nu, X_1=w,X_0=x,T_0=t_0, \mathcal{H}_{t_0-}\Big\} \nonumber\\
& \hspace{3cm} \times  q_{x,m} e^{-q_{x,m}\nu} \pi_{xw,m} \phi_{x,m}(t)d\nu.  \label{eq:der2}
\end{align}
The conditional probability in (\ref{eq:der2}) can be worked out further using (\ref{eq:main})-(\ref{eq:Phin}) and (\ref{eq:indep}) as
\begin{align}
&\mathbb{P}\Big\{X(r)=y\Big\vert \Phi(x,t,W_0)=m, -\frac{\log V_0}{q_{x,m}}=\nu, X_1=w,X_0=x, T_0=t_0,\mathcal{H}_{t_0-}\Big\} \;\;\textrm{for $w\neq x$}\nonumber\\
&\hspace{0cm}=\sum_{n=1}^{\infty} \mathbb{P}\left\{X_n=y, T_n\leq r<T_{n+1}\Big\vert \Phi(x,t,W_0)=m, X_1=w, T_1=t_1, X_0=x,T_0=t_0,\mathcal{H}_{t_0-}\right\}  \nonumber\\
&\hspace{0cm}=\sum_{n=1}^{\infty} \mathbb{P}\left\{X_n=y, T_n\leq r<T_{n+1}\Big\vert \Phi(x,t,W_0)=m, X_1=w, T_1=t_1\right\} \; \textrm{with $t_1=t+\nu$} \nonumber\\
&\hspace{0cm}=\sum_{n=1}^{\infty} \sum_{j_2,\cdots,j_{n-1}}\;\;  \idotsint\limits_{\{t_2<\cdots< t_{n-1}\}} \;\;\int_{t_n=t_1}^{r} \mathbb{P}\Big\{-\sum_{\ell=1}^M \frac{\log V_n}{q_{y,\ell}} \delta_{\ell}\big(\Phi(y,t_n,W_n)\big) > r-t_n \nonumber \\
&\hspace{6cm}\Big\vert \Phi(x,t,W_0)=m, X_n=y, T_n=t_n,\cdots, X_1=w,T_1=t_1 \Big\}  \label{eq:der3} \\
&\hspace{0.5cm}\times \mathbb{P}\Big\{X_n=y, t_1 - \sum_{k=1}^{n-1}\sum_{\ell=1}^M \frac{\log V_k}{q_{j_k,\ell}}\delta_{\ell}\big(\Phi(j_k,t_k,W_k)\big) \in d t_n  \nonumber\\
&\hspace{4cm}\Big\vert \Phi(x,t,W_0)=m,  X_{n-1}=j_{n-1}, T_{n-1} =t_{n-1},\cdots, X_1=w,T_1=t_1 \Big\} \nonumber\\
&\hspace{0.5cm}\times\mathbb{P}\Big\{X_k=j_k,T_k\in dt_k, k=2,\cdots,n-1\Big\vert \Phi(x,t,W_0)=m, X_1=w, T_1=t_1\Big\}, \;\; \textrm{with $j_1=w$.} \nonumber
\end{align}
To simplify the conditional probability (\ref{eq:der3}), let $\{X_n^{\prime}, T_n^{\prime}, V_n^{\prime}, W_n^{\prime}\}_{n\geq 0}$ be sequence of random variables having the same distribution as $\{X_n, T_n, V_n, W_n\}_{n\geq 1}$ such that $V_n=V_{n-1}^{\prime}$, $W_n=W_{n-1}^{\prime}$, $X_n=X_{n-1}^{\prime}$ and $T_n=T_{n-1}^{\prime}$. The latter implies that both $X_n^{\prime}$ and $T_n^{\prime}$ satisfy the equations (\ref{eq:Xn}) and (\ref{eq:Tn}). After replacing each $X_n, T_n, V_n$, and $W_n$ by the corresponding substituted variable in (\ref{eq:der3}),
\begin{align*}
&\mathbb{P}\Big\{X(r)=y\Big\vert \Phi(x,t,W_0)=m, X_1=w, T_1=t_1, X_0=x, T_0=t_0,\mathcal{H}_{t_0-}\Big\} \nonumber\\
&=\sum_{n=1}^{\infty} \sum_{j_2,\cdots,j_{n-1}}\;\;  \idotsint\limits_{\{t_2<\cdots< t_{n-1}\}} \;\;\int_{t_n=t_1}^{r} \mathbb{P}\Big\{-\sum_{\ell=1}^M \frac{\log V_n}{q_{y,\ell}} \delta_{\ell}\big(\Phi(y,t_n,W_n)\big) > r-t_n  \nonumber \\
&\hspace{6.5cm}\Big\vert \Phi_0=m,X_n=y, T_n=t_n,\cdots,X_1=w,T_1=t_1  \Big\} \nonumber \\
&\hspace{1cm}\times \mathbb{P}\Big\{X_n=y, t_1 - \sum_{k=1}^{n-1}\sum_{\ell=1}^M \frac{\log V_k}{q_{j_k,\ell}}\delta_{\ell}\big(\Phi(j_k,t_k,W_k)\big) \in d t_n \nonumber \\
&\hspace{3.5cm}\Big\vert \Phi_0=m, X_{n-1}=j_{n-1},\cdots,X_1=w,T_1=t_1\Big\} \nonumber\\
&\hspace{1cm}\times\mathbb{P}\Big\{X_k=j_k,T_k\in dt_k, k=2,\cdots,n-1\Big\vert \Phi_0=m, X_1=w, T_1=t_1\Big\} \\
&=\sum_{n^{\prime}=0}^{\infty} \sum_{j_1^{\prime},\cdots,j_{n^{\prime}-1}^{\prime}}\;\;  \idotsint\limits_{\{t_1^{\prime}<\cdots< t_{n^{\prime}-1}^{\prime}\}} \;\;\int_{t_{n^{\prime}}^{\prime}=t_0^{\prime}}^{r} \mathbb{P}\Big\{-\sum_{\ell=1}^M \frac{\log V_{n^{\prime}}^{\prime}}{q_{y,\ell}} \delta_{\ell}\big(\Phi(y,t_{n^{\prime}}^{\prime},W_{n^{\prime}}^{\prime})\big) > r-t_{n^{\prime}}^{\prime}  \nonumber \\
&\hspace{7cm} \Big\vert \Phi_0=m, X_{n^{\prime}}^{\prime}=y, T_{n^{\prime}}^{\prime}=t_{n^{\prime}}^{\prime}, \cdots,X_0^{\prime}=w, T_0^{\prime}=t_0^{\prime}  \Big\} \nonumber \\
&\hspace{1cm}\times \mathbb{P}\Big\{X_{n^{\prime}}^{\prime}=y, t_0^{\prime}- \sum_{k^{\prime}=0}^{n^{\prime}-1}\sum_{\ell=1}^M \frac{\log V_{k^{\prime}}^{\prime}}{q_{j_{k^{\prime}}^{\prime},\ell}}\delta_{\ell}\big(\Phi(j_{k^{\prime}}^{\prime},t_{k^{\prime}}^{\prime},W_{k^{\prime}}^{\prime})\big) \in d t_{n^{\prime}}^{\prime} \nonumber \\
&\hspace{3.5cm}\Big\vert \Phi_0=m,   X_{n^{\prime}-1}^{\prime}=j_{n^{\prime}-1}^{\prime}, T_{n^{\prime}-1}^{\prime} =t_{n^{\prime}-1}^{\prime},\cdots, X_0^{\prime}=w, T_0^{\prime}=t_0^{\prime} \Big\} \nonumber\\
&\hspace{1cm}\times\mathbb{P}\Big\{X_{k^{\prime}}^{\prime}=j_{k^{\prime}}^{\prime},T_{k^{\prime}}^{\prime}\in d t_{k^{\prime}}^{\prime}, k^{\prime}=1,\cdots,n^{\prime}-1\Big\vert \Phi_0=m, X_0^{\prime}=w, T_0^{\prime}=t_0^{\prime}\Big\}\\
&=\sum_{n^{\prime}=0}^{\infty} \sum_{j_1^{\prime},\cdots,j_{n^{\prime}-1}^{\prime}}\;\;  \idotsint\limits_{\{\tau_1^{\prime}<\cdots< \tau_{n^{\prime}-1}^{\prime}\}} \;\;\int_{\tau_{n^{\prime}}^{\prime}=0}^{r-t-\nu} \mathbb{P}\Big\{-\sum_{\ell=1}^M \frac{\log V_{n^{\prime}}^{\prime}}{q_{y,\ell}} \delta_{\ell}\big(\Phi(y,\tau_{n^{\prime}}^{\prime},W_{n^{\prime}}^{\prime})\big) > r-t-\nu-\tau_{n^{\prime}}^{\prime}  \nonumber \\
&\hspace{7cm} \Big\vert \Phi_0=m, X_{n^{\prime}}^{\prime}=y, T_{n^{\prime}}^{\prime}=\tau_{n^{\prime}}^{\prime},\cdots, X_0^{\prime}=w, T_0^{\prime}=0 \Big\} \nonumber \\
&\hspace{1cm}\times \mathbb{P}\Big\{X_{n^{\prime}}^{\prime}=y, - \sum_{k^{\prime}=0}^{n^{\prime}-1}\sum_{\ell=1}^M \frac{\log V_{k^{\prime}}^{\prime}}{q_{j_{k^{\prime}}^{\prime},\ell}}\delta_{\ell}\big(\Phi(j_{k^{\prime}}^{\prime},\tau_{k^{\prime}}^{\prime},W_{k^{\prime}}^{\prime})\big) \in d \tau_{n^{\prime}}^{\prime} \nonumber \\
&\hspace{3.5cm}\Big\vert \Phi_0=m,   X_{n^{\prime}-1}^{\prime}=j_{n^{\prime}-1}^{\prime}, T_{n^{\prime}-1}^{\prime} =\tau_{n^{\prime}-1}^{\prime},\cdots, X_0^{\prime}=w, T_0^{\prime}=0 \Big\} \nonumber\\
&\hspace{1cm}\times\mathbb{P}\Big\{X_{k^{\prime}}^{\prime}=j_{k^{\prime}}^{\prime},T_{k^{\prime}}^{\prime}\in d \tau_{k^{\prime}}^{\prime}, k^{\prime}=1,\cdots,n^{\prime}-1\Big\vert \Phi_0=m, X_0^{\prime}=w, T_0^{\prime}=0\Big\}\\
&=\sum_{n^{\prime}=0}^{\infty} \mathbb{P}\Big\{X_{n^{\prime}}^{\prime}=y,T_{n^{\prime}}^{\prime}\leq r-t -\nu < T_{n^{\prime}+1}^{\prime}\Big\vert \Phi_0=m, X_0^{\prime}=w, T_0^{\prime}=0\Big\}\\
&=\mathbb{P}\Big\{X^{\prime}(r-t-\nu)=y \Big\vert \Phi_0=m, X_0^{\prime}=w,T_0^{\prime}=0\Big\} \quad \; \textrm{by (\ref{eq:trans1})}\\
&=P_{wy}^{(m)}(r-t-\nu),
\end{align*}
with $t_0^{\prime}=t+\nu$ and $j_0^{\prime}=w$. Notice that we have used change-of-index $n^{\prime}=n-1$, $k^{\prime}=k-1$ and $j_i^{\prime}=j_{i+1}$ on the first equality, employed (\ref{eq:recTn}) on the second, and applied change-of-variable $\tau_{k^{\prime}}^{\prime}=t_{k^{\prime}}^{\prime}-t_0^{\prime}$ to the inner integral on the third equality. It is important to note that for each $k=0,1,\cdots,n$ the event $\{\Phi(j_{k},t_{k},W_{k})=m\}$ under the probability measure $\mathbb{P}\{\bullet\big\vert X_{k}=j_{k},T_{k}=t_{k},\cdots,X_0=x_0,T_0=t_0\}$ is equivalent with $\{\Phi(j_{k},\tau_{k},W_{k})=m\}$ under $\mathbb{P}\{\bullet\big\vert X_{k}=j_{k},T_{k}=\tau_{k},\cdots,X_0=x_0,T_0=0\}$. This is due to the fact that in the absence of censoring, time shifting the epoch times $\{T_k\}_{k\geq1}$ by $t_0^{\prime}$ does not alter the likelihood function of the observed sample paths, see (\ref{eq:likelihood}). This observation was taken into account to arrive at the fourth equality in the above series of equalities. Following (\ref{eq:der2}),
\begin{align}
&\mathbb{P}\{X(r)=y, t\leq T_1\leq r \big\vert X(t)=x,\mathcal{H}_{t-}\} \nonumber\\
&\hspace{2cm}=\sum_{m=1}^M \phi_{x,m}(t) q_{x,m}\int_0^{r-t} e^{-q_{x,m}\nu} \sum_{w\neq x} \pi_{xw,m} P_{wy}^{(m)}(r-t-\nu) d\nu  \nonumber\\
&\hspace{2cm}=\sum_{m=1}^M \phi_{x,m}(t) q_{x,m} e^{-q_{x,m}(r-t)}\int_0^{r-t} e^{q_{x,m}u} \sum_{w\neq x} \pi_{xw,m} P_{wy}^{(m)}(u) du, \label{eq:buktiI}
\end{align}
where the last equality was obtained after a change-of-variable $u=r-t-\nu$. Given that $T_1$ is the first jump time of $X$ since the process starts at $X_0=X(T_0)=x$ at time $T_0=t>0$, we have
\begin{align}
\mathbb{P}\{X(r)=y, T_1\geq r \big\vert X(t)=x,\mathcal{H}_{t-}\}=&\mathbb{P}\{T_1\geq r \big\vert X_0=x, T_0=t,\mathcal{H}_{t_0-}\}\delta_x(y)  \nonumber\\
&\hspace{-5cm}=\mathbb{P}\Big\{t-\sum_{\ell=1}^M \delta_{\ell}\big(\Phi(x,t,W_0)\big) \frac{\log V_0}{q_{x,\ell}} \geq r\Big\vert X_0=x, T_0=t, \mathcal{H}_{t_0-}\Big\}\delta_x(y)   \nonumber\\
&\hspace{-5cm}=\sum_{m=1}^M \mathbb{P}\Big\{t-\sum_{\ell=1}^M \delta_{\ell}\big(\Phi(x,t,W_0)\big) \frac{\log V_0}{q_{x,\ell}} \geq r, \Phi(x,t,W_0)=m\Big\vert X_0=x, T_0=t, \mathcal{H}_{t_0-}\Big\}\delta_x(y)   \nonumber\\
&\hspace{-5cm}=\sum_{m=1}^M \mathbb{P}\Big\{-\frac{\log V_0}{q_{x,m}} \geq r-t, \Phi(x,t,W_0)=m\Big\vert X_0=x, T_0=t,\mathcal{H}_{t_0-}\Big\}\delta_x(y)   \nonumber\\
&\hspace{-5cm}=\sum_{m=1}^M \mathbb{P}\Big\{-\frac{\log V_0}{q_{x,m}} \geq r-t\Big\}\mathbb{P}\Big\{\Phi(x,t,W_0)=m\Big\vert X_0=x, T_0=t, \mathcal{H}_{t_0-}\Big\}\delta_x(y)  \nonumber\\
&\hspace{-5cm}=\sum_{m=1}^M e^{-q_{x,m}(r-t)} \phi_{x,m}(t)\delta_x(y). \label{eq:buktiII}
\end{align}

Thus, by the law of total probability, we obtain following (\ref{eq:buktiI}) and (\ref{eq:buktiII})
\begin{align*}
\mathbb{P}\big\{X(r)=y \big\vert X(t)=x,\mathcal{H}_{t-}\big\} =&\mathbb{P}\big\{X(r)=y, T_1\geq r \big\vert X(t)=x,\mathcal{H}_{t-}\big\} \\&\hspace{2cm}+\mathbb{P}\big\{X(r)=y, t\leq T_1\leq r \big\vert X(t)=x,\mathcal{H}_{t-}\big\} \\
&\hspace{-3cm}=\sum_{m=1}^M \phi_{x,m}(t) \Big( e^{-q_{x,m}(r-t)}\delta_x(y) +q_{x,m} e^{-q_{x,m}(r-t)}\int_0^{r-t} e^{q_{x,m}\nu} \sum_{w\neq x} \pi_{xw,m} P_{wy}^{(m)}(\nu) d\nu   \Big)\\
&\hspace{-3cm}=\sum_{m=1}^M \phi_{x,m}(t) P_{xy}^{(m)}(r-t), 
\end{align*}
where on the last equality we used (\ref{eq:intdiff}). Hence our assertion of the theorem is established. $\blacksquare$

\begin{cor}
Since the uniform random variable $W_0$ is independent of $X$ and $\mathbb{P}\{\Phi(x,t,W_0)=m\vert X(t)=x,\mathcal{H}_{t-}\}=\phi_{x,m}(t)$ with $\sum_{m=1}^M \phi_{x,m}(t)=1$ for $x\in\mathbb{S}$ and $t\geq 0$, we have 
\begin{align*}
\mathbb{P}\big\{X(r)=y\big\vert \Phi(x,t,W_0)=m, X(t)=x, \mathcal{H}_{t-}\big\}=P_{xy}^{(m)}(r-t).
\end{align*}
The above identity dictates that $X$ (\ref{eq:main}) is indeed a conditional Markov jump process.
\end{cor}


\section{Proof of Theorem \ref{theo:Fisher}}\label{sec:Fisher}

\noindent \textbf{Proof.}
Following the likelihood function (\ref{eq:likelihood3}) and the identity (\ref{eq:id10}), we have
\begin{align*}
\frac{\partial}{\partial \theta_j}\log f_{\theta}(\mathcal{X})=&\sum_{k=1}^K  \frac{\partial}{\partial \theta_j}\log f_{\theta}(X^k) 
=\sum_{k=1}^K \mathbb{E}_{\theta}\Big(\frac{\partial}{\partial \theta_j} \log f_{\theta}(X^k,\Phi^k)\Big\vert X^k\Big)\\
=&\sum_{k=1}^K \sum_{m=1}^M \Big(\frac{\partial}{\partial \theta_j} \log f_{\theta}(X^k,\Phi^k=m) \Big)f_{\theta}(\Phi^k=m\big\vert X^k).
\end{align*}
Therefore, differentiating both sides of the above equality w.r.t variable $\theta_i$ gives
\begin{align}\label{eq:derv1}
\frac{\partial^2}{\partial \theta_i \partial \theta_j}\log f_{\theta}(\mathcal{X})
=&\sum_{k=1}^K \sum_{m=1}^M \Big(\frac{\partial^2}{\partial \theta_i\partial \theta_j} \log f_{\theta}(X^k,\Phi^k=m) \Big)f_{\theta}(\Phi^k=m\big\vert X^k) \nonumber\\
&\hspace{0cm}+ \sum_{k=1}^K \sum_{m=1}^M \Big(\frac{\partial}{\partial \theta_j} \log f_{\theta}(X^k,\Phi^k=m) \Big)\frac{\partial}{\partial \theta_i}f_{\theta}(\Phi^k=m\big\vert X^k).
\end{align}
After some computations using Bayes' formula and the identity (\ref{eq:id10}) one can show that
\begin{align*}
\frac{\partial }{\partial \theta_i} f_{\theta}(\Phi^k=m\big\vert X^k)
=&f_{\theta}(\Phi^k=m\big\vert X^k)\frac{\partial}{\partial\theta_i}\log f_{\theta}(X^k,\Phi^k=m) \nonumber\\
&\hspace{0cm}-f_{\theta}(\Phi^k=m\big\vert X^k)\mathbb{E}_{\theta}\Big(\frac{\partial}{\partial \theta_i}\log f_{\theta}(X^k,\Phi^k)\Big\vert X^k\Big).
\end{align*}
Replacing the derivative $\frac{\partial }{\partial \theta_i} f_{\theta}(\Phi^k=m\big\vert X^k)$ in the equation (\ref{eq:derv1}) with the one above yields
\begin{align*}
\frac{\partial^2}{\partial \theta_i \partial \theta_j}\log f_{\theta}(\mathcal{X})=&\sum_{k=1}^K \sum_{m=1}^M \Big(\frac{\partial^2}{\partial \theta_i\partial \theta_j} \log f_{\theta}(X^k,\Phi^k=m) \Big)f_{\theta}(\Phi^k=m\big\vert X^k) \nonumber\\
&\hspace{-2cm}+ \sum_{k=1}^K \sum_{m=1}^M \Big(\frac{\partial}{\partial \theta_j} \log f_{\theta}(X^k,\Phi^k=m) \Big)\Big[f_{\theta}(\Phi^k=m\big\vert X^k)\frac{\partial}{\partial\theta_i}\log f_{\theta}(X^k,\Phi^k=m) \nonumber\\
&\hspace{1cm}-f_{\theta}(\Phi^k=m\big\vert X^k)\mathbb{E}_{\theta}\Big(\frac{\partial}{\partial \theta_i}\log f_{\theta}(X^k,\Phi^k)\Big\vert X^k\Big)  \Big],
\end{align*}
leading to (\ref{eq:InfMatr}) after rearranging the sum. This ends the assertion of the proposition. $\blacksquare$

\subsection{Equivalence of the information matrix (\ref{eq:InfMatr}) with Louis' general formula (1982)}\label{sec:subsecC}
Starting from the observed information matrix (\ref{eq:InfMatr}), we obtain after adding and subtracting the term $\Big[\sum_{k=1}^K \mathbb{E}_{\theta}\Big(\frac{\partial}{\partial \theta_i}\log f_{\theta}(X^k,\Phi^k)\Big\vert X^k\Big)\Big]\Big[\sum_{\ell=1}^K \mathbb{E}_{\theta}\Big(\frac{\partial}{\partial \theta_j}\log f_{\theta}(X^{\ell},\Phi^{\ell})\Big\vert X^{\ell}\Big)\Big]$ in the identity (\ref{eq:InfMatr}),
\begin{align}
\frac{\partial^2}{\partial \theta_i \partial \theta_j}\log f_{\theta}(\mathcal{X}) 
=&\sum_{k=1}^K \mathbb{E}_{\theta}\Big(\frac{\partial^2}{\partial \theta_i \partial \theta_j} \log f_{\theta}(X^k,\Phi^k)\Big\vert X^k\Big) \nonumber\\
&\hspace{0cm}+\sum_{k=1}^K \mathbb{E}_{\theta}\Big( \Big[\frac{\partial}{\partial \theta_i}\log f_{\theta}(X^k,\Phi^k)\Big]
\Big[\frac{\partial}{\partial \theta_j}\log f_{\theta}(X^k,\Phi^k)\Big]\Big\vert X^k\Big) \nonumber\\
&\hspace{0cm}+\Bigg\{\Big[\sum_{k=1}^K \mathbb{E}_{\theta}\Big(\frac{\partial}{\partial \theta_i}\log f_{\theta}(X^k,\Phi^k)\Big\vert X^k\Big)\Big]\Big[\sum_{\ell=1}^K \mathbb{E}_{\theta}\Big(\frac{\partial}{\partial \theta_j}\log f_{\theta}(X^{\ell},\Phi^{\ell})\Big\vert X^{\ell}\Big)\Big]  \nonumber\\
&\hspace{1cm}-\sum_{k=1}^K\mathbb{E}_{\theta}\Big( \frac{\partial}{\partial \theta_i}\log f_{\theta}(X^k,\Phi^k)\Big\vert X^k\Big)\mathbb{E}_{\theta}\Big( \frac{\partial}{\partial \theta_j}\log f_{\theta}(X^k,\Phi^k)\Big\vert X^k\Big)\Bigg\} \nonumber\\
&-\Big[\sum_{k=1}^K \mathbb{E}_{\theta}\Big(\frac{\partial}{\partial \theta_i}\log f_{\theta}(X^k,\Phi^k)\Big\vert X^k\Big)\Big]\Big[\sum_{\ell=1}^K \mathbb{E}_{\theta}\Big(\frac{\partial}{\partial \theta_j}\log f_{\theta}(X^{\ell},\Phi^{\ell})\Big\vert X^{\ell}\Big)\Big] \nonumber\\
=& \sum_{k=1}^K \mathbb{E}_{\theta}\Big(\frac{\partial^2}{\partial \theta_i \partial \theta_j} \log f_{\theta}(X^k,\Phi^k)\Big\vert \mathcal{X}\Big) \nonumber\\
&\hspace{0cm}+\Bigg\{\sum_{k=1}^K \mathbb{E}_{\theta}\Big( \Big[\frac{\partial}{\partial \theta_i}\log f_{\theta}(X^k,\Phi^k)\Big]
\Big[\frac{\partial}{\partial \theta_j}\log f_{\theta}(X^k,\Phi^k)\Big]\Big\vert \mathcal{X}\Big)  \nonumber\\ 
&\hspace{1cm}+\sum_{k=1}^K\sum_{\ell\neq k}^K \mathbb{E}_{\theta}\Big(\Big[\frac{\partial}{\partial \theta_i}\log f_{\theta}(X^k,\Phi^k)\Big]\Big[\frac{\partial}{\partial \theta_j}\log f_{\theta}(X^{\ell},\Phi^{\ell})\Big]\Big\vert \mathcal{X}\Big)\Bigg\} \nonumber\\
&-\Big[\sum_{k=1}^K \mathbb{E}_{\theta}\Big(\frac{\partial}{\partial \theta_i}\log f_{\theta}(X^k,\Phi^k)\Big\vert \mathcal{X}\Big)\Big]\Big[\sum_{\ell=1}^K \mathbb{E}_{\theta}\Big(\frac{\partial}{\partial \theta_j}\log f_{\theta}(X^{\ell},\Phi^{\ell})\Big\vert \mathcal{X}\Big)\Big]  \nonumber\\
=&  \mathbb{E}_{\theta}\Big(\sum_{k=1}^K\frac{\partial^2}{\partial \theta_i \partial \theta_j} \log f_{\theta}(X^k,\Phi^k)\Big\vert \mathcal{X}\Big) \nonumber\\
&+\mathbb{E}_{\theta}\Big(\Big[\sum_{k=1}^K \frac{\partial}{\partial \theta_i}\log f_{\theta}(X^k,\Phi^k)\Big] \Big[\sum_{\ell=1}^K \frac{\partial}{\partial \theta_j}\log f_{\theta}(X^{\ell},\Phi^{\ell})\Big]\Big\vert \mathcal{X}\Big)  \label{eq:InfMatr2} \\
&-\Big[ \mathbb{E}_{\theta}\Big(\sum_{k=1}^K\frac{\partial}{\partial \theta_i}\log f_{\theta}(X^k,\Phi^k)\Big\vert \mathcal{X}\Big)\Big]\Big[\mathbb{E}_{\theta}\Big(\sum_{\ell=1}^K \frac{\partial}{\partial \theta_j}\log f_{\theta}(X^{\ell},\Phi^{\ell})\Big\vert \mathcal{X}\Big)\Big],  \nonumber
\end{align} 
where the second equality was due to independence of the sample paths $(X^k,\Phi^k)$ for $k=1,\cdots, K$. By denoting $\log L_c(\theta)=\sum_{k=1}^K \log f_{\theta}(X^k,\Phi^k)$ and taking account of linearity of the partial derivative operators, the last equality corresponds to the general matrix formula of Louis \cite{Louis}. $\blacksquare$


\section{Proof of Proposition \ref{prop:prop3}}

We give the proof for $J_p(\widehat{\phi}_{y,n},\widehat{\phi}_{x,m})$ and leave the rest to the reader, or refer to the proof of Proposition 2 in Frydman and Surya \cite{Frydman2020} which was based on the Louis' general matrix formula \cite{Louis}. Following the log-likelihood function (\ref{eq:loglike}), we have for any $(x,y)\in \mathbb{S}$ and $1\leq m,n\leq M$ that
\begin{align*}
\frac{\partial}{\partial \phi_{x,m}}\log f_{\theta}(X^k,\Phi^k)=&\frac{\Phi_{k,m}B_x^k}{\phi_{x,m}} - B_x^k,\\
\frac{\partial^2}{\partial \phi_{x,m} \partial \phi_{y,n}}\log f_{\theta}(X^k,\Phi^k)=&-\frac{\Phi_{k,n}B_x^k}{\phi_{x,m}\phi_{y,n}}\delta_m(n)\delta_x(y),
\end{align*}
from which it follows on the fact that the information $B_x^k$ contains in the path $X^k$, i.e., $\in X^k$,
\begin{align*}
\mathbb{E}_{\theta}\left( \frac{\partial}{\partial \phi_{x,m}}\log f_{\theta}(X^k,\Phi^k)\Big\vert X^k   \right)=&\frac{\widehat{\Phi}_{k,m}(\theta)B_x^k}{\phi_{x,m}} - B_x^k,\\
\mathbb{E}_{\theta}\left(\frac{\partial^2}{\partial \phi_{x,m} \partial \phi_{y,n}}\log f_{\theta}(X^k,\Phi^k)\Big\vert X^k\right)=&-\frac{\widehat{\Phi}_{k,n}(\theta)B_x^k}{\phi_{x,m}\phi_{y,n}}\delta_m(n)\delta_x(y).
\end{align*}
On account that $\Phi_{k,n}\Phi_{k,m}=\delta_m(n)\Phi_{k,n}$ and $B_x^k B_y^k=\delta_x(y)B_y^k$, we obtain after some calculations
\begin{align*}
&\left( \frac{\partial}{\partial \phi_{x,m}}\log f_{\theta}(X^k,\Phi^k)  \right)
\left( \frac{\partial}{\partial \phi_{y,n}}\log f_{\theta}(X^k,\Phi^k)  \right)\\
&\hspace{2cm}=
\frac{\delta_m(n)\delta_x(y)}{\phi_{x,m}\phi_{y,n}}\Phi_{k,n}B_y^k -\frac{\delta_x(y)}{\phi_{x,m}}\Phi_{k,m}B_y^k-\frac{\delta_x(y)}{\phi_{y,n}}\Phi_{k,n}B_y^k + \delta_x(y) B_y^k.
\end{align*}
Define for any $x\in\mathbb{S}$ and $1\leq m\leq M$, $\widehat{B}_{x,m}(\theta)=\sum_{k=1}^K \widehat{\Phi}_{k,m}(\theta)B_x^k.$ After some computations,
\begin{align*}
&\sum_{k=1}^K \mathbb{E}_{\theta}\left[\left( \frac{\partial}{\partial \phi_{x,m}}\log f_{\theta}(X^k,\Phi^k)  \right)
\left( \frac{\partial}{\partial \phi_{y,n}}\log f_{\theta}(X^k,\Phi^k)  \right) \Big\vert X^k \right]\\
&\hspace{2cm}=
\frac{\delta_m(n)\delta_x(y)}{\phi_{x,m}\phi_{y,n}}\widehat{B}_{y,n}(\theta) -\frac{\delta_x(y)}{\phi_{x,m}}\widehat{B}_{y,m}(\theta)-\frac{\delta_x(y)}{\phi_{y,n}}\widehat{B}_{y,n}(\theta) + \delta_x(y) B_y^k.
\end{align*}
Putting the sum of conditional expectations all together, the claim is established from (\ref{eq:InfMatr}). $\blacksquare$

\section{Proof of Proposition \ref{prop:MatIc}}

From the log-likelihood function (\ref{eq:loglike}) and applying the results of Proposition C1 in \cite{Frydman2020}, 
\begin{align*}
I_c(\phi_{x^{\prime},n},q_{xy,m})=&\mathbb{E}\left[ \Big(\frac{\partial \log f_{\theta}(X^k,\Phi^k)}{\partial \phi_{x^{\prime},n}}\Big) \Big(\frac{\partial \log f_{\theta}(X^k,\Phi^k)}{\partial q_{xy,m}}\Big)  \right]\\
=&\frac{1}{\phi_{x^{\prime},n}q_{xy,m}}\mathbb{E}\left[\Big(\Phi_{k,n}B_{x^{\prime}}^k  - \phi_{x^{\prime},n} B_{x^{\prime}}^k \Big)\Big( \Phi_{k,m}N_{xy}^k - q_{xy,m} \Phi_{k,m}T_x^k   \Big)\right]\\
=&\frac{1}{\phi_{x^{\prime},n}q_{xy,m}}\Big[   \delta_m(n)\mathbb{E}\Big(\Phi_{k,n}B_{x^{\prime}}^k N_{xy}^k\Big) -\delta_m(n)q_{xy,m}\mathbb{E}\Big(\Phi_{k,n}B_{x^{\prime}}^k T_x^k\Big)\\
&\hspace{2cm}-\phi_{x^{\prime},n}\mathbb{E}\Big(\Phi_{k,m}B_{x^{\prime}}^k N_{xy}^k\Big) + \phi_{x^{\prime},n}q_{xy,m}\mathbb{E}\Big(\Phi_{k,m}B_{x^{\prime}}^k T_x^k\Big)\Big]\\
=&0,
\end{align*}
for $x,x^{\prime},y\in \mathbb{S}$ and $1\leq n,m\leq M$ for which $\phi_{x^{\prime},n}$ and $q_{xy,m}$ are all non zero. It follows that $I_c(\theta)$ is a $(Mp^2\times Mp^2)$block diagonal matrix, the same structure as the covariance matrix $\Sigma(\theta_0)$. $\blacksquare$

It remains to derive the sub-matrices $I_c(q_{x^{\prime}y^{\prime},n},q_{xy,m})$ and $I_c(\phi_{x^{\prime},n},\phi_{x,m}).$ First, recall that
\begin{align*}
I_c(q_{x^{\prime}y^{\prime},n},q_{xy,m})=\;&\mathbb{E}\left[ \Big(\frac{\partial \log f_{\theta}(X^k,\Phi^k)}{\partial q_{x^{\prime}y^{\prime},n}}\Big) \Big(\frac{\partial \log f_{\theta}(X^k,\Phi^k)}{\partial q_{xy,m}}\Big) \right]\\
=&\frac{1}{ q_{x^{\prime}y^{\prime},n} q_{xy,m} } \mathbb{E}\left[\Big(\Phi_{k,n} N_{x^{\prime}y^{\prime}}^k- q_{x^{\prime}y^{\prime},n} \Phi_{k,n}T_{x^{\prime}}^k\Big) \Big(\Phi_{k,m}N_{xy}^k-q_{xy,m}\Phi_{k,m}T_{x}^k\Big)  \right]\\
=&\frac{\delta_m(n)}{q_{x^{\prime}y^{\prime},n}q_{xy,m}}\Big[ \mathbb{E}\Big(\Phi_{k,n}N_{x^{\prime}y^{\prime}}^k N_{xy}^k\Big)-q_{xy,m}\mathbb{E}\Big(\Phi_{k,n}N_{x^{\prime}y^{\prime}}^k T_x^k\Big) \\
&\hspace{2.5cm}-q_{x^{\prime}y^{\prime},n}\mathbb{E}\Big(\Phi_{k,n} N_{xy}^k T_{x^{\prime}}^k\Big)   
+q_{xy,m}q_{x^{\prime}y^{\prime},n}\mathbb{E}\Big(\Phi_{k,n}T_x^k T_{x^{\prime}}^k\Big) \Big],
\end{align*}
which by applying the results of Proposition C1 in \cite{Frydman2020}, it simplifies into %
\begin{eqnarray*}\label{eq:Jqq}
I_c(q_{x^{\prime}y^{\prime},n},q_{xy,m})=\frac{\delta_m(n)\delta_x(x^{\prime})\delta_y(y^{\prime})}{q_{x^{\prime}y^{\prime},n}}\alpha^{\top}S_n\int_0^T e^{Q_n u} e_x du. \quad \blacksquare
\end{eqnarray*}

After some calculations similar to those above, we have
\begin{align*}
I_c(\phi_{y,n},\phi_{x,m})=\mathbb{E}\left[ \Big(\frac{\partial \log f_{\theta}(X^k,\Phi^k)}{\partial \phi_{y,n}}\Big) \Big(\frac{\partial \log f_{\theta}(X^k,\Phi^k)}{\partial \phi_{x,m}}\Big) \right]
=\frac{\alpha_{y}}{\phi_{y,m}}\delta_x(y)\Big(\delta_m(n)-\phi_{x,m}\Big).\quad \blacksquare
\end{align*}

\section{Proof of Theorem \ref{theo:maintheo}}

The proof of consistency for the MLE $\widehat{\theta}_0$ (\ref{eq:theMLE}) follows from the convergence of $L_K(\theta)$ to $L(\theta)$ and the fact that $\widehat{\theta}_0$ solves the equation $L_K^{\prime}(\theta_0)=0$ and that $\theta_0$ is the global and unique solution of $L^{\prime}(\theta)=0$, see \cite{Frydman2020} for details. As the convergence of $L_K(\theta)$ to $L(\theta)$ occurs with probability one, $\widehat{\theta}_0\stackrel{\mathbb{P}}{\Longrightarrow} \theta_0$ as $K\rightarrow \infty$. To show the asymptotic normality of the MLE, let $\ell_K(\theta):=\frac{1}{K}\log  f_{\theta}(\mathcal{X})$. Since $\ell_K(\theta)$ is continuously differentiable for all $\theta\in\Theta$, see the log-likelihood function (\ref{eq:likelihood3}), we have by the regularity condition (\ref{eq:regular}) and an application of the Mean Value Theorem,
\begin{align*}
\ell_K^{\prime}(\theta_0)=\ell_K^{\prime}(\widehat{\theta}_0) -\ell_K^{\prime\prime}(\theta_1)\big(\widehat{\theta}_0-\theta_0), \quad \textrm{for \;$\theta_0\leq \theta_1\leq \widehat{\theta}_0$},
\end{align*}
from which it follows on account of $\ell_K^{\prime}(\widehat{\theta}_0)=0$ that 
\begin{align}\label{eq:clt}
\sqrt{K}\big(\widehat{\theta}_0-\theta_0\big)=-\big[\ell_K^{\prime\prime}(\theta_1)\big]^{-1}\sqrt{K} \ell_K^{\prime}(\theta_0).
\end{align}
By consistency of the MLE $\widehat{\theta}_0$, $\theta_1\rightarrow \theta_0$ as $K$ increases. Hence, by independence of the sample paths $\{X^k\}$, $\ell_K^{\prime\prime}(\theta_1)\rightarrow -I_p(\theta_0)$ with probability one as $K$ increases. By identity (\ref{eq:id10}) and the regularity condition (\ref{eq:bounded}), $\ell_K^{\prime}(\theta_0)=L_K^{\prime}(\theta_0)$. Furthermore, since $\theta_0$ is the global maximum of $L(\theta)$, i.e., $L^{\prime}(\theta_0)=0$,
it turns out that (\ref{eq:clt}) has the same asymptotic distribution as the random variable
\begin{align*}
I_p^{-1}(\theta_0)\sqrt{K}\big(L_K^{\prime}(\theta_0)-L^{\prime}(\theta_0)\big).
\end{align*}
See Ch. 2 of van der Vaart \cite{deVaart}. By the multivariate central limit theorem, the above has asymptotic multivariate normal distribution with mean zero and covariance matrix $I_p^{-1}(\theta_0)I_c(\theta_0) I_p^{-1}(\theta_0).$ $\blacksquare$

\end{document}